\documentclass[review]{elsarticle}
\usepackage[a4paper, margin=1in]{geometry}
\usepackage{lineno,hyperref}
\usepackage{hyperref}
\usepackage{amsmath}
 \usepackage{graphicx}
\graphicspath{{images/}}
\usepackage[]{caption}
\usepackage[]{subcaption}
\usepackage{listings}
\usepackage{multirow}
\usepackage{float}
\usepackage{color}

\modulolinenumbers[100]

\journal{Journal of Computational Science}

\begin{document}

\begin{frontmatter}

\title{A parallel fluid solid coupling model using \textit{LAMMPS} and \textit{Palabos} based on the immersed boundary method}

%% Group authors per affiliation:
%\author{Jifu Tan\fnref{myfootnote}}
%\address{Radarweg 29, Amsterdam}
%\fntext[myfootnote]{Since 1880.}

%% or include affiliations in footnotes:
\author[mymainaddress]{Jifu Tan}
\author[mymainaddress2]{Talid Sinno}
\author[mymainaddress2]{Scott L Diamond\corref{cor2}}
%\ead[url]{www.elsevier.com}

%\author[mysecondaryaddress]{Global Customer Service\corref{mycorrespondingauthor}}
%\cortext[cor1]{Corresponding author, Email address: sld@seas.upenn.edu}
\cortext[cor2]{Corresponding author, Email address: jifutan@niu.edu, sld@seas.upenn.edu}
%\ead{support@elsevier.com}

\address[mymainaddress]{Department of Mechanical Engineering\\ Northern Illinois University\\ DeKalb, IL 60115, USA}
\address[mymainaddress2]{Department of Chemical and Biomolecular Engineering\\ University of Pennsylvania\\ Philadelphia, PA 19104, USA}
%\address[mysecondaryaddress]{360 Park Avenue South, New York}

\begin{abstract}
The study of viscous fluid flow coupled with rigid or deformable solids has many applications in biological and engineering problems, e.g., blood cell transport, drug delivery, and particulate flow. We developed a partitioned approach to solve this coupled Multiphysics problem.  The fluid motion was solved by \textit{Palabos} (Parallel Lattice Boltzmann Solver), while the solid displacement and deformation was simulated by \textit{LAMMPS} (Large-scale Atomic/Molecular Massively Parallel Simulator). The coupling was achieved through the immersed boundary method (IBM). The code modeled both rigid and deformable solids exposed to flow. The code was validated with the Jeffery orbits of an ellipsoid particle in shear flow, red blood cell stretching test, and effective blood viscosity flowing in tubes. It demonstrated essentially linear scaling from 512 to 8192 cores for both strong and weak scaling cases. The computing time for the coupling increased with the solid fraction. An example of the fluid-solid coupling was given for flexible filaments (drug carriers) transport in a flowing blood cell suspensions, highlighting the advantages and capabilities of the developed code. 
\end{abstract}


\end{frontmatter}

\linenumbers

\section{Introduction}
Fluid flows containing solid particles are common in engineering and medicine, e.g., suspension flows \cite{nott1994pressure}, sedimentation\cite{feng1994direct}, cell transport in blood flow \cite{fedosov2010blood,vahid2015microparticle,doddi2008lateral,sinha2016shape,tan2012influence}, and platelet deposition on blood vessel walls \cite{pivkin2006blood,fogelson2008immersed,lu2016multiscale}. The dynamic behavior in such phenomena is complex due to interactions between individual particles as well as interactions between the particles and the surrounding fluid and bounding walls. Moreover, the presence of highly deformable particles, such as blood cells, vesicles and polymers, make it particularly challenging to accurately describe the dynamics in such systems. Understanding the interactions between the particulate components and the fluid is essential for optimized particulate design or detailed particulate flow behavior prediction, e.g., enhance particle mixing\cite{tsuji1993discrete}, fluid coking\cite{gao2008cfd}, drug carrier design\cite{muller2016understanding, vahid2015microparticle, radhakri2013temporal}, cell separation\cite{kruger2014deformability}, and blood clotting\cite{lu2016multiscale,yazdani2017general,fogelson2015fluid}.

The present paper is focused on blood flows containing a high concentration of deformable cells that are similar in size to the vessel diameter, thus requiring explicit consideration of the particle mechanics. Readers interested in the modeling of other classes of particulate flows, e.g., those relevant to industrial applications, are referred to other studies, e.g., Ref. \cite{zhu2008discrete, kloss2012models, chiesa2005numerical}. Numerous methods have been developed to model blood flows containing cells such as red blood cells and platelets. The boundary element method is an example of a very efficient technique for these types of flows\cite{freund2013flow, pozrikidis2001effect,zhao2012shear} as it formulates boundary value problems as boundary integral equations. Thus, it only requires discretization of the surface rather than the volume. However, the boundary element method requires explicit knowledge of a fundamental or analytical solution of the differential equations, e.g., linear partial differential equations. Thus, it is limited to Stokes flow conditions. The Arbitrary-Lagrangian-Eulerian method (ALE) is another approach that has been widely used to model fluid-solid interactions\cite{donea1982arbitrary, souli2000ale}. In ALE methods the fluid mesh boundary conforms to the solid boundaries on the interface, i.e., the nodes at the fluid-solid interfaces are shared. However, the ALE technique is computationally very expensive because repeated mesh generation is necessary for flows where particles experience large deformations (e.g., for red blood cells). 

By contrast, non-conforming mesh methods eliminate the mesh regeneration step. In these methods, an Eulerian mesh is used for the fluid and a Lagrangian mesh is used for the solid. These two meshes are independent, i.e., they do not share nodes across the interfaces. In this case, boundary conditions are applied through imposed constraints on the interfaces. The Immersed Boundary Method (IBM) is a popular example of a non-conforming mesh technique\cite{peskin2002immersed,mittal2005immersed}. Here, the velocity and force boundary conditions at the interfaces are imposed though interpolation functions that transfer velocities and forces across the domains. Consequently, the IBM approach may be regarded as an interface between two essentially separate simulators--one for evolving the particulate phase and the other for evolving the fluid. As we will show in this paper, the decoupled nature of the two components leads to a high degree of versatility in the context of software development. More generally, the present work represents an example of an emerging paradigm in multi-physics/multiscale modeling in which multiple existing (and independent) software packages are connected by a relatively simple interface to generate new functionality. Examples in the literature of such approaches include software packages for general fluid-solid coupling\cite{constant2016implementation}, sedimentation\cite{sun2016sedifoam}, atomic-continuum coupling\cite{cosden2013hybrid}, and fluid flow coupled with the discrete element method\cite{kloss2012models}. 
    
Another important aspect related to the implementation of methods for solving particulate flows is the portability to high performance computing (HPC) platforms so that large system sizes and long simulation times relevant to the phenomena of interest are accessible. An example of a reported HPC-enabled blood flow simulation is the work of \cite{peters2010multiscale}, in which blood flow simulations in patient-specific coronary arteries at spatial resolutions ranging from the centimeter scale down to 10 $\mu m$ were performed on 294,912 cores. In Ref.\cite{clausen2010parallel}, 2,500 deformable red blood cells in suspension under flow were simulated on IBM Blue Gene/P supercomputers. Additional examples include large-scale simulations of blood flow in the heart \cite{griffith2012immersed,shadden2010computational,shadden2015lagrangian}, cell separation in microfluidic flows\cite{rossinelli2015silico,kruger2014deformability}, and blood flow in the brain\cite{perdikaris2016multiscale}. However, there are only a few parallel open sourced fluid solid coupling codes, such as the immersed boundary (IB) method with support for adaptive mesh refinement (AMR) \textit{IBAMR}\cite{griffith2007adaptive}, the vascular flow simulation tool \textit{SimVascular}\cite{taylor1998finite,updegrove2016simvascular}, the CFDEM project using computational fluid dynamics and discrete element methods\cite{seil2017lbdemcoupling, kloss2012models}. A general open source fluid solid coupling tool with versatile functionalities (e.g., deformable solids) and independent fluid and solid simulator is still missing. 

In order to overcome challenges related to programming, model standardization, dissemination and sharing, and efficient implementation on parallel computing architectures, here we introduce a simple but effective implementation of the immersed boundary method for simulating deformable particulate flows using popular open source software. This project was also inspired by the previous work from LBDEM\cite{seil2017lbdemcoupling, seil2017onset}. The fluid solver is based on the lattice Boltzmann method(LBM), chosen because of its efficient parallelization across multiple processor environments\cite{peng2008parallel,clausen2010parallel}. LBM is a versatile fluid flow solver engine and has been used to model diverse situations, such as flows in porous media with complex geometries\cite{boek2010lattice,spaid1997lattice,yu2017hemodynamic} and multiphase flows\cite{he1999lattice,shan1993lattice}. The solid particles in the fluids, particularly, the deformable blood cells are modeled as a coarse grained cell membrane model using particle based solid solver. Many LBM based fluid solid coupling work can be found in literature\cite{reasor2012coupling,zavodszky2017cellular,kruger2014deformability,hyakutake2015numerical,shi2013numerical}, some of them haven't demonstrated the large scale parallel performance\cite{hyakutake2015numerical,shi2013numerical}. Further, none of them are open sourced yet. In the present paper we employ a general LBM fluid solver \textit{Palabos}\cite{latt2007hydrodynamic, latt2009palabos} and couple it with the \textit{LAMMPS} software package \cite{plimpton1995fast} for describing particle dynamics. Even \textit{LAMMPS} provides a LBM fluid solver through USER-LB extension\cite{ollila2011fluctuating,mackay2013hydrodynamic}, the functionality of the fluid solver is limited, e.g., it can only model fluid with simple geometries and apply boundary velocities on the z direction.  Thus, it is still necessary to build a large-scale parallel simulation tool for fluid-solid coupling based on widely used, general purpose open source codes. The implementation of a stable, efficient large-scale parallel solver for either fluid flow or particle dynamics is nontrivial, requiring training in scientific computing and software engineering. Both \textit{Palabos} and \textit{LAMMPS} are efficiently parallelized and clearly demonstrated by ample documentation and actively supported from large on-line communities:\textit{LAMMPS} has an active mailing list that has hundreds questions and answers posted daily. These codes are also extensively validated by numerous examples and publications. Moreover, user-developers may easily extend the functionality of either package by implementing additional features, e.g., interaction potentials, integrators, fluid models, etc. 

To the best of our knowledge, this is the first time \textit{Palabos} has been coupled with \textit{LAMMPS} in an immersed boundary method framework. The remainder of the paper is structured as follows. First, short introductions are provided in Section \ref{sec:mtd} describing the fluid solver (\ref{ssec:lbm}), the solid solver (\ref{ssec:lmp}), the immersed boundary method (\ref{ssec:ibm}), and the spatial decomposition for the coupling (\ref{ssec:sdc}). Next in Section \ref{rst:validation}, a validation test and convergence study are presented for a single ellipsoid in a shear flow. The parallel performance of the IBM solver is studied in Section \ref{rst:performance}. Finally, an example of flexible filament transport in blood cell suspensions is described in Section \ref{rst:app}, highlighting the advantages and capabilities of the new solver. Conclusions and discussions are provided in Section \ref{con}. 

\section{Methods}\label{sec:mtd}
\subsection{Lattice Boltzmann fluid solver: \textit{Palabos}}\label{ssec:lbm}
\textit{Palabos} is an open source computational fluid dynamics solver based on the lattice Boltzmann method (LBM). It is designed in C++ with parallel features using Message Passing Interface (MPI). It has been employed widely in both academic and industrial settings. The LBM has been used extensively in blood flow modeling \cite{zhang2008red,kruger2011efficient,macmeccan2009simulating,lu2016multiscale,tan2016characterization,yu2017hemodynamic}. Reviews of the underlying theory for the LBM can be found in the literature \cite{chen1998lattice,he1997lattice,succi2001lattice,ladd1994numericalTheory}. LBM is usually considered as a second-order accurate method in space and time \cite{inamuro1997accuracy}. The fundamental quantity underpinning the LBM is the density distribution function $f_i(\boldsymbol{x},t)$ in phase space $(\boldsymbol{x}, \vec{\boldsymbol{c}}_{i})$, where $t$ denotes the time and $\vec{\boldsymbol{c}}_{i}$ denotes the lattice velocity. The evolution of the density distribution function involves streaming and collision processes,
\begin{equation}\label{eqn:lbm}
\underbrace{f_i(\boldsymbol{x}+\vec{\boldsymbol{c}}_{i},t+1)-f_i(\boldsymbol{x},t)}_\text{streaming}=\underbrace{\frac{1}{\tau}(f_i^{eq}-f_i(\boldsymbol{x},t))}_\text{BGK collision}+F_i,
\end{equation}
where the simplest Bhatnagar$-$Gross$-$Krook (BGK) scheme \cite{qian1992lattice} is used for the collision term and $F_i$ is the body force term that will be used to represent immersed cell boundaries\cite{guo2002discrete}.
The equilibrium distribution$f_i^{eq}$ is given by 
\begin{equation}\label{eqn:BGKequilibrium}
f_i^{eq}=w_i\rho\left(1+\frac{\vec{\boldsymbol{c}}_i\cdot\vec{\boldsymbol{u}}}{c_s^2}+\frac{(\vec{\boldsymbol{c}}_i\cdot\vec{\boldsymbol{u}})^2}{2c_s^4}-\frac{\vec{\boldsymbol{u}}^2}{2c_s^2}\right),
\end{equation}
where $w_i$ are the weight coefficients and $c_s$ is the speed of sound. The density $\rho$ and velocity $\vec{\boldsymbol{u}}$ may be calculated as
\begin{equation}\label{eqn:rho_u}
\rho = \sum f_i, \hspace{2mm}\rho\vec{\boldsymbol{u}}=\sum f_i\vec{\boldsymbol{c}}_i+\frac{1}{2}\vec{\boldsymbol{g}},
\end{equation}
where $\vec{\boldsymbol{g}}$ is the external force density vector that is related to $F_i$ as
\begin{equation}\label{eqn:BGKequilibrium}
F_i=(1-\frac{1}{2\tau})w_i\left(\frac{\vec{\boldsymbol{c}}_i-\vec{\boldsymbol{u}}}{c_s^2}+\frac{\vec{\boldsymbol{c}}_i\cdot\vec{\boldsymbol{u}}}{c_s^4}\vec{\boldsymbol{c}}_i\right)\cdot\vec{\boldsymbol{g}}.
\end{equation}
see Ref.\cite{guo2002discrete} for more details on force terms. The fluid viscosity $\nu$ is related to the relaxation parameter $\tau$ as
\begin{equation}\label{eqn:viscosity}
\nu=c_s^2(\tau-0.5)=\frac{\tau-0.5}{3}.
\end{equation}
In all LBM simulations reported in this paper, the fluid domain is discretized using a uniform D3Q19 lattice; see ref.\cite{succi2001lattice}. The fluid density $\rho$, $\vec{\boldsymbol{u}}$ and density distribution function $f_i$ are initialized at the equilibrium distribution calculated from Eqn.\ref{eqn:BGKequilibrium} based on the initial fluid velocity. During each time step, streaming and collision steps are performed on $f_i$ according to Eqn.\ref{eqn:lbm}. Specifically, the $f_i$ is translated in the direction of the discretized velocity vector $\vec{\boldsymbol{c}}_{i}$ during the streaming step; then, $f_i$ is updated based on the equilibrium distribution, $f_i^{eq}$, the relaxation parameter, $\tau$, and the force density, $F_i$, which is passed to the LBM from the immersed solid objects, e.g., blood cells.   

\subsection{Particle based solid solver: \textit{LAMMPS} for deformable cells and particles}\label{ssec:lmp}
\textit{LAMMPS} was originally designed as a molecular dynamics simulation tool\cite{plimpton1995fast}. In molecular dynamics, a potential function is defined to model the interactions between atoms. The force on each atom is calculated as the derivative of the potential with respect to the atomic coordinates and the atomic system motion is updated based on numerical integrations of Newton's 2nd law of motion. The \textit{LAMMPS} package has now been extended to include a variety of additional dynamical engines, including peridynamics\cite{parks2008implementing,silling2000reformulation}, smooth particle hydrodynamics\cite{monaghan1983shock,monaghan1992smoothed}, dissipative particle dynamics\cite{groot1997dissipative,afshar2013exploiting}, and stochastic rotation dynamics\cite{ihle2001stochastic}. LIGGGHTS(\textit{LAMMPS} improved for general granular and granular heat transfer simulations) is also an extension of \textit{LAMMPS} for discrete element method particle simulation\cite{kloss2012models}. Many predefined potentials, functions, and ODE integrators in \textit{LAMMPS} make it extremely powerful for modeling atomic, soft matter\cite{genheden2015simple}, and biological systems\cite{fedosov2010multiscale,pivkin2006blood}. 

\begin{figure}[H]
\centering
\begin{subfigure}{.5\textwidth}
  \centering
  \includegraphics[width=0.7\linewidth]{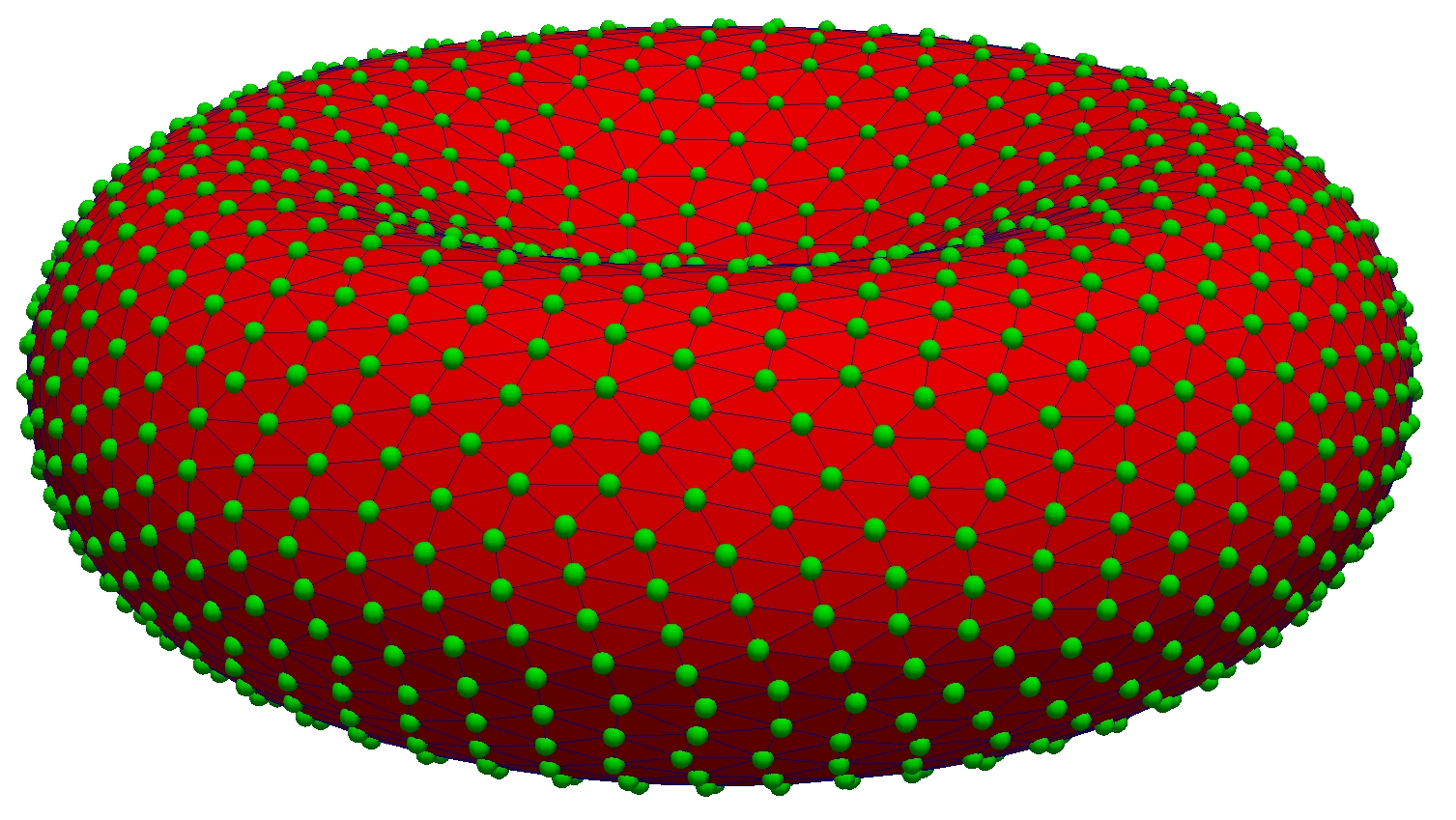}
  \caption{A particle based coarse grained cell membrane model}
  \label{fig:cgcm}
\end{subfigure}%
\begin{subfigure}{.5\textwidth}
  \centering
  \includegraphics[width=0.94\linewidth]{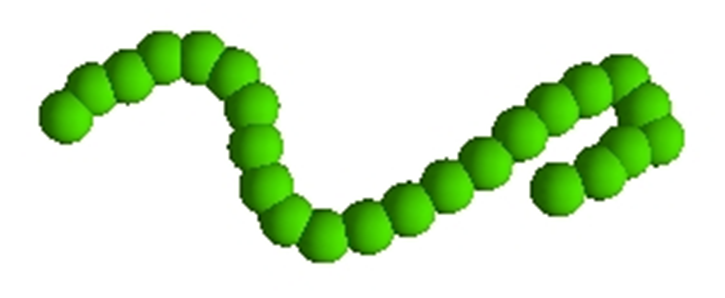}
  \caption{A polymer chain model}
  \label{fig:plmr}
\end{subfigure}
\caption{Models for a deformable red blood cell and a polymer chain implemented in \textit{LAMMPS}: (a) a particle based coarse grained red blood cell membrane model that can bear stretching and bending. The particles on the cell membrane are interacting with potentials. (b) a particle connected polymer chain model with stretching and bending resistance.}
\label{fig:rbcplmr}
\end{figure}

A coarse-grained membrane model consisting of many interacting particles\cite{pivkin2008accurate, reasor2012coupling, dao2006molecularly,tan2017numerical,yu2017hemodynamic} is used to simulate red blood cells, as shown in Fig.\ref{fig:cgcm}. The membrane model can bear stretching and bending. Constraints to maintain the constant membrane surface area and enclosed cell volume are imposed through harmonic potentials. The viscosity ratio of cytoplasm over blood plasma is about 5. Here we treat the blood cell internal and external viscosity to be the same for saving computational cost. Readers interested in different viscosity models are referred to Ref.\cite{doddi2008lateral,reasor2013determination,freund2011cellular}.
The potential function for a red blood cell (RBC) used in the current work is given by
\begin{equation}\label{eqn:membrane-potential}
U(\boldsymbol{X}_i)=U_{stretch}+U_{bending}+U_{area}+U_{volume},
\end{equation}
where the stretching energy $U_{stretch}$ is used to represent the cytoskeleton's resistance to deformation. The bending energy $U_{bending}$ represents the rigidity of the membrane bilayer imparted to the cytoskeleton. The last two terms are the constraints for maintaining constant membrane surface area and cell volume.
The stretching potential is given by: 
\begin{equation}\label{eqn:inPlanePotential}
U_{stretch}=\sum_{j\in 1...N_s}\left[\frac{k_BTl_m}{4p}\frac{3x_j^2-2x_j^3}{1-x_j}+\dfrac{k_p}{l_j}\right],
\end{equation}
where $l_m$ is the maximum bond length, the $j$th bond length ratio is $x_j=l_j/l_m$. $l_m$ was set to be 2 times the equilibrium bond length. $N_s$ is the number of springs, $p$ is the persistence length, $k_B$ is the Boltzmann constant, $T$ is the temperature, and $k_p$ is the repulsive potential constant. Once $p$ is specified, $k_p$ may be found using the value of $x_j=0.5$ at the equilibrium where the net force is zero.

The bending energy is defined as
\begin{equation}\label{eqn:bendingEnergy}
U_{bending}=\sum_{j=1...N_s}k_b(1-cos(\theta_j-\theta_0)),
\end{equation}
where $k_b$ is the bending constant and $\theta_j$ is the instantaneous angle formed by the two outward surface norms of two adjacent triangular meshes that share the same edge $j$. $\theta_0$ is the corresponding equilibrium, or spontaneous, angle. 

Constraints for constant membrane surface area and cell volume are imposed though area/volume dependent harmonic potentials,
\begin{equation}\label{eqn:areaEnergy}
U_{area}=\dfrac{k_g(A-A_0)^2}{2A_0}+\sum_{j=1...N_t}\dfrac{k_l(A_j-A_{j0})^2}{2A_{j0}},
\end{equation}
\begin{equation}\label{eqn:volumeEnergy}
U_{volume}=\dfrac{k_v(V-V_0)^2}{2V_0},
\end{equation}
where $k_g, k_l$ are the global and local area constraint constants, $N_t$ is the number of triangular surfaces, $A, A_0$ are the instantaneous and spontaneous total surface area of the cell membrane, $A_j, A_{j0}$ are the instantaneous and spontaneous surface area for the $j$th triangle surface, $k_v$ is the volume constraint constant, and $V, V_0$ are the instantaneous and the equilibrium cell volume. 

The parameters used in the coarse-grained membrane model can be related to membrane properties used in continuum model\cite{dao2006molecularly,fedosov2010multiscale}, e.g., the shear modulus, $\mu_0$, which is given by
\begin{equation}\label{eqn:shearModulus}
\mu_0=\dfrac{\sqrt{3}k_BT}{4pl_mx_0}\left[\dfrac{x_0}{2(1-x_0)^3}-\dfrac{1}{4(1-x_0)^2}+\dfrac{1}{4}\right]+\dfrac{3\sqrt{3}k_p}{4l_0^3}.
\end{equation}
where $x_0=l_0/l_m$ and $l_0$ is the bond length at equilibrium. Eqn.\ref{eqn:shearModulus} holds assuming the membrane surface area is preserved during simulation, thus the contribution from $U_{area}$ can be safely ignored. Interested readers can refer to Ref.\cite{dao2006molecularly,fedosov2010multiscale} for details. 

For polymer particles, as shown in Fig.\ref{fig:plmr}, the stretching energy is the same as Eqn.(\ref{eqn:inPlanePotential}), while the bending energy is a harmonic function of the angle deviation, 
\begin{equation}\label{eqn:bendingEnergyWorm}
U_{bending}^{p}=k_b^p(\theta_j-\theta_0)^2,
\end{equation}
where the superscript $p$ refers to polymer and the other variables are defined as in Eqn.(\ref{eqn:bendingEnergy}). 

To avoid the overlapping of the particles from different solid objects, e.g, cells, polymers, etc. a Morse potential was used for inter-particle interaction,
\begin{equation}\label{eqn:morsePotential}
U_{morse}=D_0[e^{-2\alpha(r-r_0)}-2e^{-\alpha(r-r_0)}], \quad r<r_c.
\end{equation}
where $D_0$ is the energy scale, $\alpha$ controls the width of the potential, $r$ is the distance between particles from different solid objects, $r_0$ is the equilibrium distance, $r_c$ is the cutoff distance. 

All the parameters used in the simulation are listed in Table \ref{tab:paraPtlCellMix}. 

\subsection{Fluid-solid coupling: the immersed boundary method}\label{ssec:ibm}
\begin{figure}[H]
\centering
\includegraphics[width=0.6\textwidth]{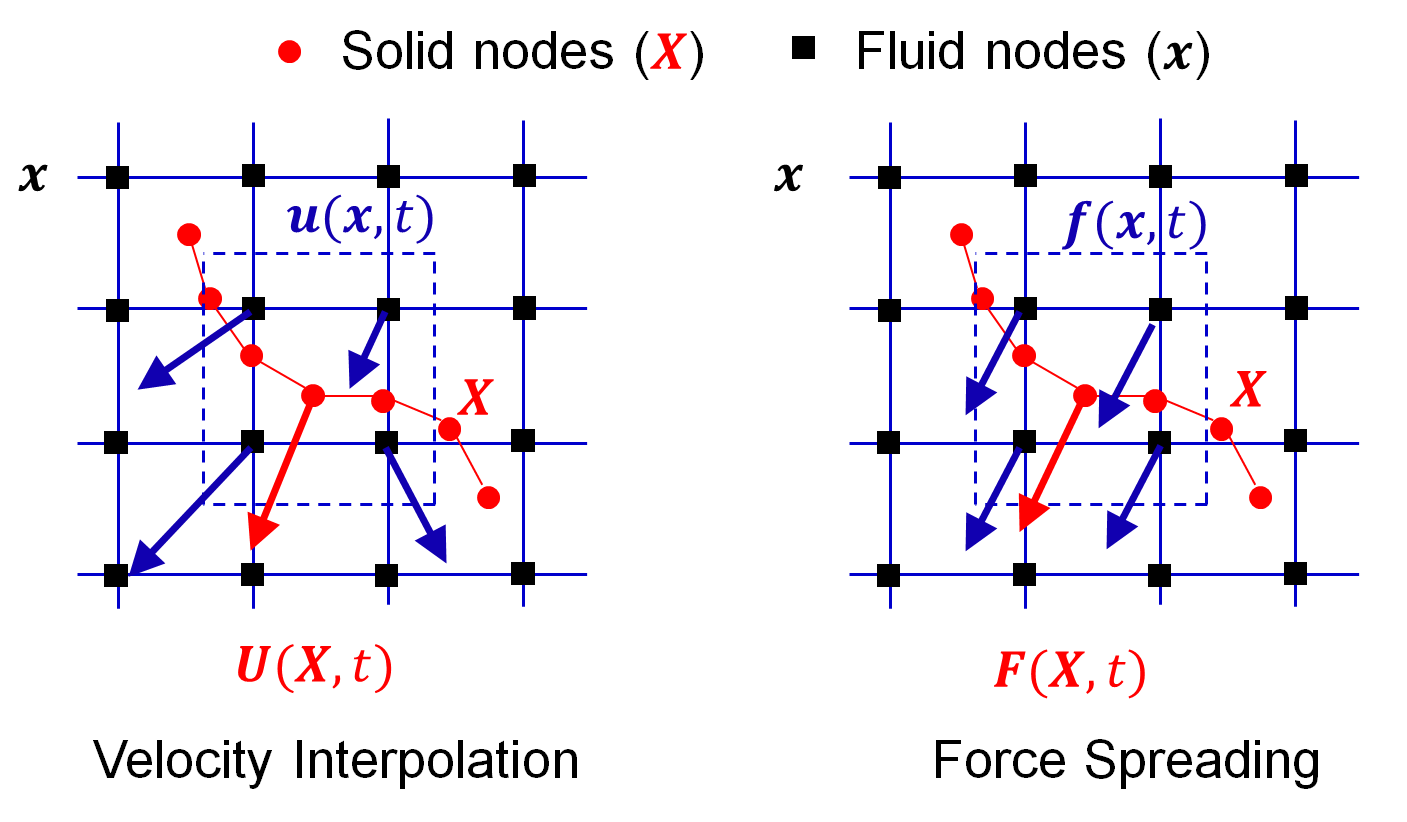}
\caption{An illustration of the fluid-solid coupling through the immersed boundary method. The solid velocity for each particle, $\boldsymbol{U}(\boldsymbol{X},t)$, is interpolated from the local fluid velocities, $\boldsymbol{u}(\boldsymbol{x},t)$, while the solid force at each particle $\boldsymbol{F}(\boldsymbol{X},t)$ is spread out onto the local fluid nodes, $\boldsymbol{f}(\boldsymbol{x},t)$. The influence range of the central solid particle on fluid is shown in dashed rectangles.}
\label{fig:ibmInterpolationSpreading}
\end{figure}

The immersed boundary method (IBM) was used to model the coupling between fluid and solid. The combination of LBM and IBM was first used to model fluid-particle interaction problems in Ref. \cite{feng2004immersed}. Details of the immersed boundary method formulation may be found in Refs.  \cite{peskin2002immersed,mittal2005immersed,feng2004immersed}. Briefly, the solid velocity at each particle position is obtained through velocity interpolation from local fluid nodes, while fluid forces are obtained by spreading the local solid forces. Specifically, for an immersed solid with coordinates $\boldsymbol{X}$, the velocity $\boldsymbol{U}(\boldsymbol{X},t)$ is interpolated from the local fluid velocity $\boldsymbol{u}(\boldsymbol{x},t)$, while the solid force, $\boldsymbol{F}(\boldsymbol{X},t)$, calculated from Eqn.\ref{eqn:membrane-potential} is spread out into the local fluid grid points as a force density $\boldsymbol{f}(\boldsymbol{x},t)$:
\begin{equation}\label{eqn:ibmVelocityInterpret}
\boldsymbol{U}(\boldsymbol{X},t)=\int\boldsymbol{u}(\boldsymbol{x},t)\delta(\boldsymbol{x}-\boldsymbol{X})d\boldsymbol{x},
\end{equation}
\begin{equation}\label{eqn:ibmForceSpread}
\boldsymbol{f}(\boldsymbol{x},t)=\int\boldsymbol{F}(\boldsymbol{X},t)\delta(\boldsymbol{x}-\boldsymbol{X})d\boldsymbol{X},
\end{equation}

where $\delta(\boldsymbol{x})$ is the delta function. In a typical numerical implementation in three dimensions, $\delta(\boldsymbol{x})$ is constructed as the product of one-dimensional functions, i.e., $\delta(\boldsymbol{x})=\phi(x)\phi(y)\phi(z)$ with $\phi(r)$ defined as
\begin{equation}\label{eqn:deltaPhiSimple}
\phi(r)=
\begin{cases}
& 0,\quad\quad \text{otherwise}\\
& \dfrac{1}{4}\left(1+cos\left(\dfrac{\pi r}{2}\right)\right), \quad\quad |r|\leq 2,
\end{cases}
\end{equation}
where $r$ is the distance between solid particles and fluid nodes. Different choices of the interpolation function influence the coupling accuracy, the influence range of the solid particles on the fluid, and the computational cost \cite{liu2012properties,yang2009smoothing}. Schematic representations of the velocity interpolation and force spreading process used in the present implementation are shown in Fig.\ref{fig:ibmInterpolationSpreading}. 

The coupling strategy described above becomes numerically unstable for rigid objects. In such situations, a different fluid-solid interaction (FSI) approach must be applied \cite{goldstein1993modeling,wu2010simulating}. In the present IBM formulation for rigid particles, the FSI force on each particle is used to assemble the total force and torque on the whole rigid body. The translation and rotation of the rigid object are then updated based on Newton's 2nd law of motion. The FSI force can be expressed as:
\begin{equation}\label{eqn:FSIRigid}
f^{FSI}(\boldsymbol{X},t)=\rho_f(\boldsymbol{u}(\boldsymbol{X},t)-\boldsymbol{U}(\boldsymbol{X},t))/\delta t,
\end{equation}
where $\rho_f$ is the fluid density, $\boldsymbol{u}(\boldsymbol{X},t)$ is the fluid velocity at the solid boundary node $\boldsymbol{X}$, which has to be obtained through interpolation by Eqn.(\ref{eqn:ibmVelocityInterpret}). $\boldsymbol{U}(\boldsymbol{X},t)$ is the solid nodal velocity. 
The idea behind Eqn.(\ref{eqn:FSIRigid}) is that the local fluid particles with incoming velocity $\boldsymbol{u}(\boldsymbol{X},t)$ will collide with the solid boundary with outgoing velocity $\boldsymbol{U}(\boldsymbol{X},t)$. The FSI force is the change of the momentum divided by the collision time $\delta t$. Other approaches have been proposed to model rigid objects in IBM, e.g., a virtual boundary formulation was used in Ref.\cite{lai2000immersed}. 

Another challenge in modeling fluid mediated solid transport is to correctly capture the fluid flow in the lubrication layer between two approaching solids or between one solid and a wall. For example, when the gap between the cell membrane and the wall is very small, e.g., the gap is smaller than a lattice space, the LBM fluid solver can not resolve the fluid flow in the thin layer. One approach is to use a finer mesh for the whole fluid domain or refine the mesh near the boundary layers\cite{mavriplis1990multigrid,hughes2005isogeometric,griffith2010parallel}. This approach would increase the LBM simulation cost as more lattice space is used. It also requires some efforts to handle the density distribution passage over the interface if a multi-grid is used in LBM. Another approach is to introduce an lubrication force to repel the cell membrane so that there are enough fluid within the gap. The lubrication repelling force was introduced to the lattice Boltzmann method in Ref\cite{ladd1997sedimentation,ladd2001lattice}. Following their work, the lubrication force derived from the lubrication theory between two identical spheres is 
\begin{equation}\label{eqn:lubricationSphere}
\boldsymbol{F}_{ij}^{lub}=-\dfrac{3\pi\nu r}{s}\hat{\boldsymbol{x}}_{ij}\hat{\boldsymbol{x}}_{ij}\cdot(\boldsymbol{u}_i-\boldsymbol{u}_j)
\end{equation}
where $r$ is the spherical radius, $\nu$ is the fluid dynamic viscosity, $s$ is dimensionless gap $s=R/r-2$ where $R$ is the central distance between two spheres. $\boldsymbol{x}_{ij}$ is the position vector difference between sphere $i$ and $j$, defined as $\boldsymbol{x}_{ij}=\boldsymbol{x}_{i}-\boldsymbol{x}_{j}$, $\hat{\boldsymbol{x}}_{ij}$ is the unit vector. $\boldsymbol{u}$ is the spherical velocity. Eqn.\ref {eqn:lubricationSphere} can also be extended to the case where a sphere approach a stationary wall by setting $\boldsymbol{u}_j=0$. Interested readers can find more details there. In the present paper, we resolve the lubrication effect using smaller fluid mesh size, e.g., the fluid mesh size was set to 100 nm when a blood cell squeezing though a 5 $\mu$m tube, as shown in section \ref{rst:bloodViscosity}. 

\subsection{Spatial decomposition for fluid-solid coupling}\label{ssec:sdc}
\begin{figure}[H]
\centering
\includegraphics[width=0.8\textwidth]{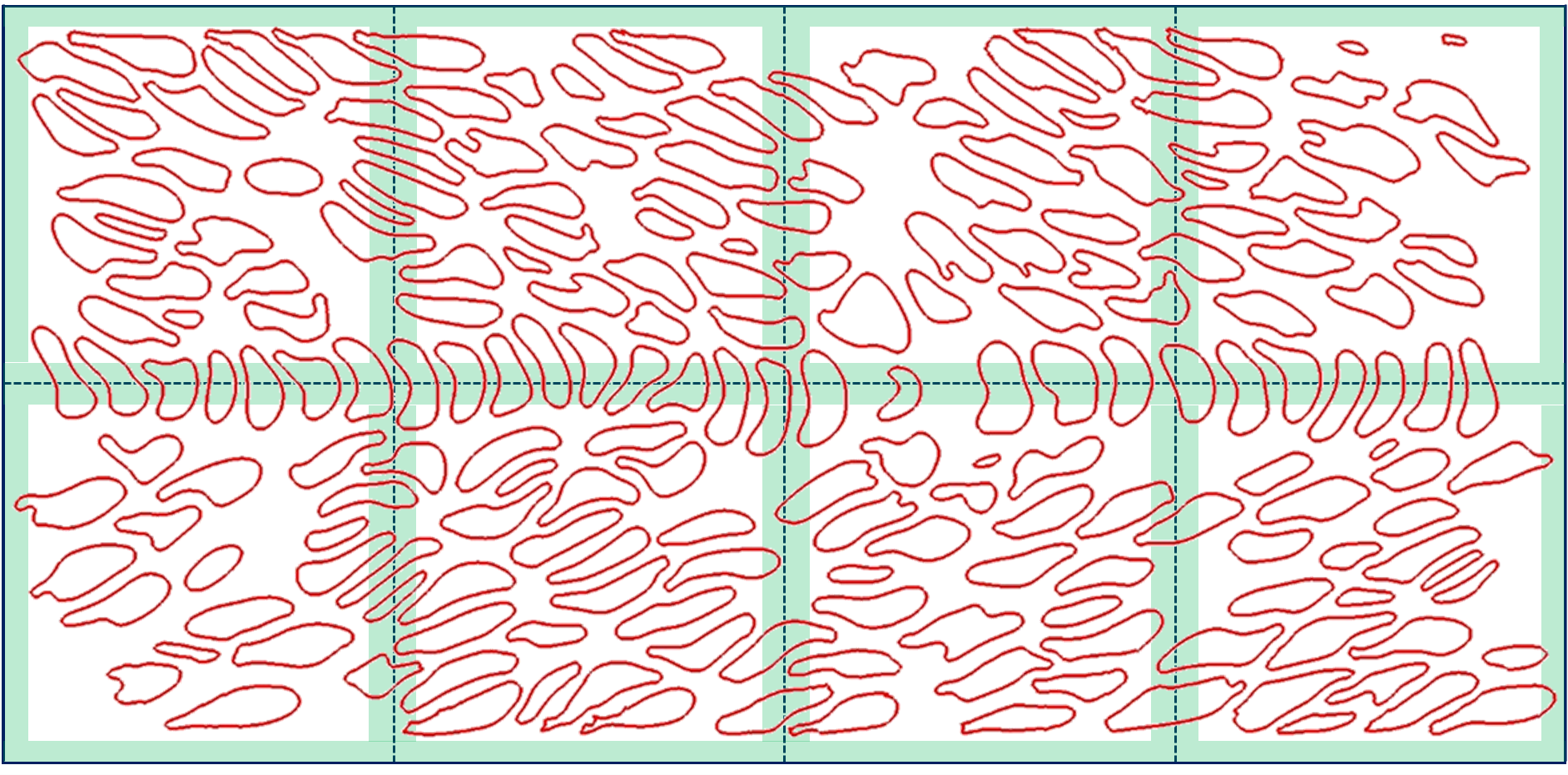}
\caption{Spatial decomposition for parallel computing for fluid solid coupling. The blood flow with cells (red lines) was divided into multiple regions. One processor was assigned to each region to calculate the both fluid and solid motions within the region, thus, the coupling between fluid and solid within the region is local to the processor. On the interface between processors, ghost layers (green stripe) were used for communication for both fluid and solid independently. An example of 8 processors were used for the task partition in the figure.}
\label{fig:geomPartition}
\end{figure}

Spatial decomposition is adopted by both \textit{Palabos} and \textit{LAMMPS} for parallel computing. In our fluid-solid coupling approach, the same spatial decomposition was applied to both the fluid and the solid so that the coupling can be effectively handled by the same compute core for a given region. Consistent spatial decomposition for the two domains ensures that individual processors have access to both the fluid grid points and solid particles within the same sub-domain. An illustration of the partitioning process is shown in Fig.\ref{fig:geomPartition}, where the whole fluid-solid system is partitioned onto 8 compute cores. Ghost layers near the boundaries of each sub-domain are used for communication between neighboring cores. This approach may not be optimal for a system where solid particles are highly heterogeneously distributed across the entire domain. However, for most cases of interest the distribution of solid particles is quite homogeneous, and in any case, the majority of the computation is dedicated to the fluid solver. The issue of load optimization for heterogeneous systems is deferred to future work. 

The FSI coupling algorithm consists almost entirely of two routines to carry out velocity interpolation and force spreading functions. As mentioned above the fluid solver usually represents the bulk of the computational demand because the solid fraction is usually quite small. Based on this intrisic asymmetry, we employed \textit{Palabos} as the driving code, while  \textit{LAMMPS} was called as an external library. To fully access all the members(e.g., particle positions, velocities, forces, etc.) from \textit{LAMMPS}, a pointer to \textit{LAMMPS} was used and passed to the two interpolation functions. An outline of the functions implemented in our IBM algorithm is shown in List\ref{lst:label}.
%\lstset{language=C++, frame=single,showspaces=false, basicstyle=\small\footnotesize}
\renewcommand\lstlistingname{List}
\begin{lstlisting}[language=C++,frame=single,showspaces=false, basicstyle=\footnotesize, caption=function declarations for IBM, label={lst:label}]
template<typename T,template<typename U> class Descriptor>
void interpolateVelocity3D(MultiBlockLattice3D<T,Descriptor> 
      &lattice,LAMMPSWrapper &wrapper);

template<typename T,template<typename U> class Descriptor>
void spreadForce3D(MultiBlockLattice3D<T,Descriptor> 
     &lattice,LAMMPSWrapper &wrapper);
\end{lstlisting}
where $lattice$ is the structure used in \textit{Palabos} to store the population distribution functions for the LBM, while $wrapper$ is a pointer to an instance of \textit{LAMMPS}. The advantage of the IBM in terms of software development is that only two functions are needed to couple a fluid solver to a solid solver. Our IBM implementation therefore only requires a few hundred lines of code beyond what is implicitly contained in the \textit{Palabos} and \textit{LAMMPS} packages. Detailed implementations can be found at https://github.com/TJFord/palabos-lammps.

\section{Results}\label{rst}
\subsection{Validation I: Ellipsoid in shear flow}\label{rst:validation}
\begin{figure}[H]
\centering
\includegraphics[width=0.4\textwidth]{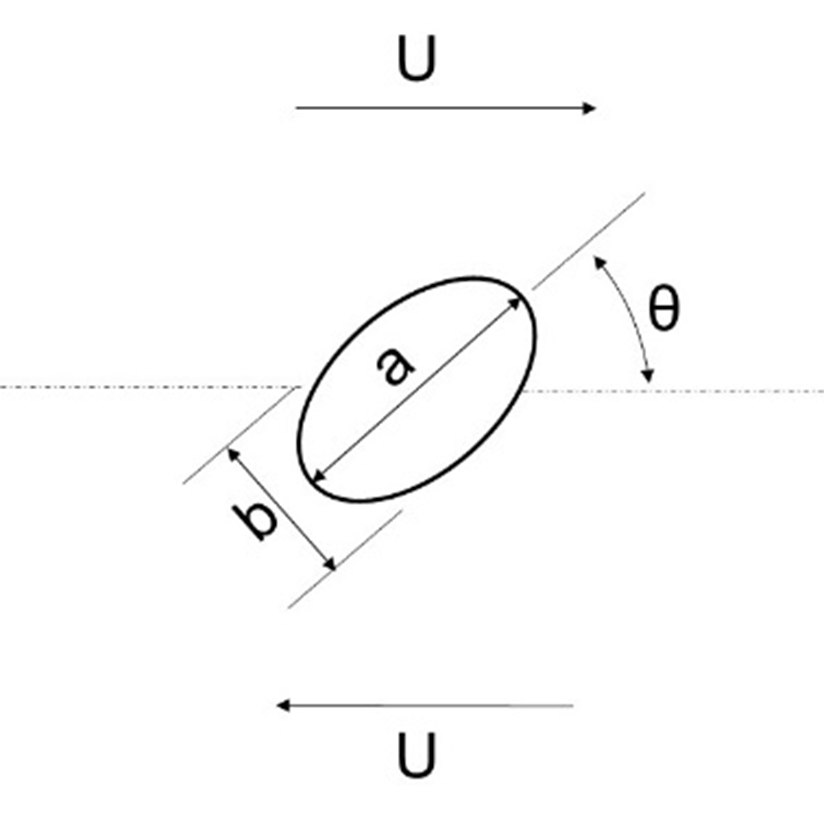}
\caption{An illustration of an ellipsoid in a shear flow. $a, b$ are the major and minor semi-axis lengths, and $U$ is the magnitude of the velocities applied at the top and bottom plates. The distance between the top and bottom parallel plate is $H$, giving a shear rate $\gamma = 2U/H$.}
\label{fig:ellipsoidInShear}
\end{figure}

\begin{figure}[H]
\centering
\includegraphics[width=0.6\textwidth]{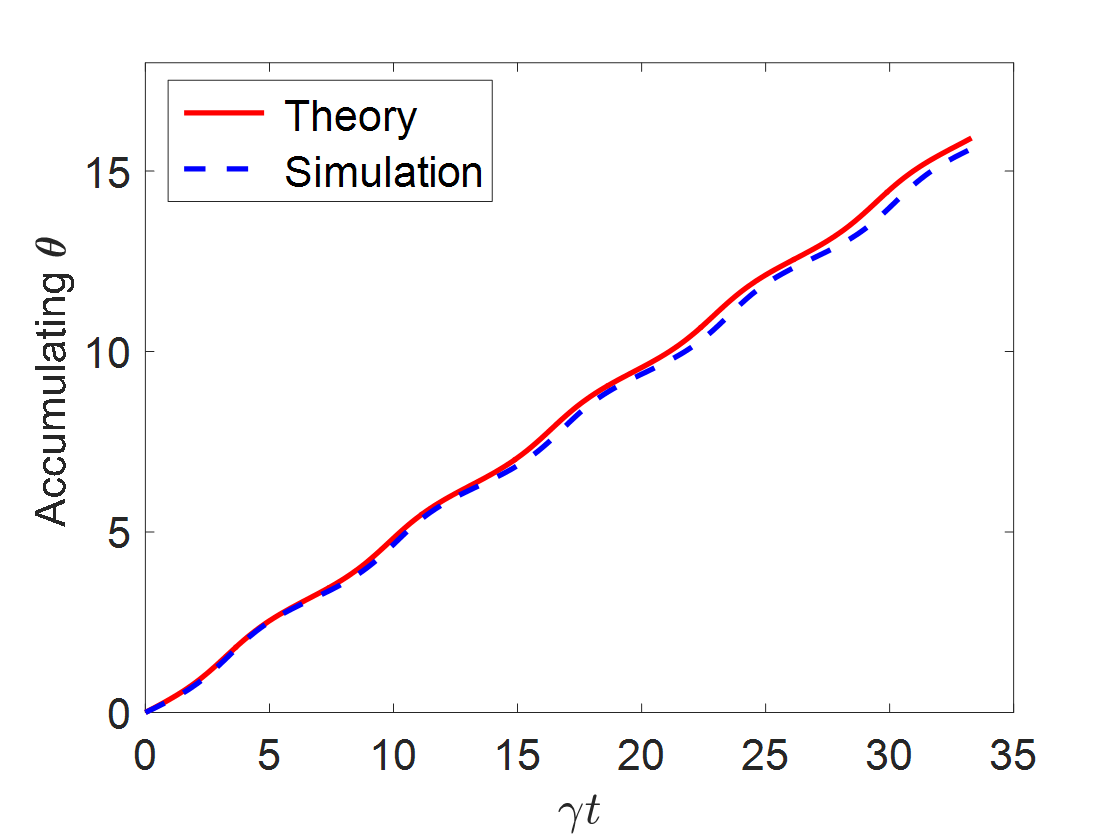}
\caption{The comparison between Jeffery's orbit given by Eqn.\ref{eqn:jefferyOrbit} and numerical results simulated by the fluid solid coupling code}
\label{fig:JefferyOrbitValidation}
\end{figure}

\begin{figure}[H]
\centering
\includegraphics[width=0.6\textwidth]{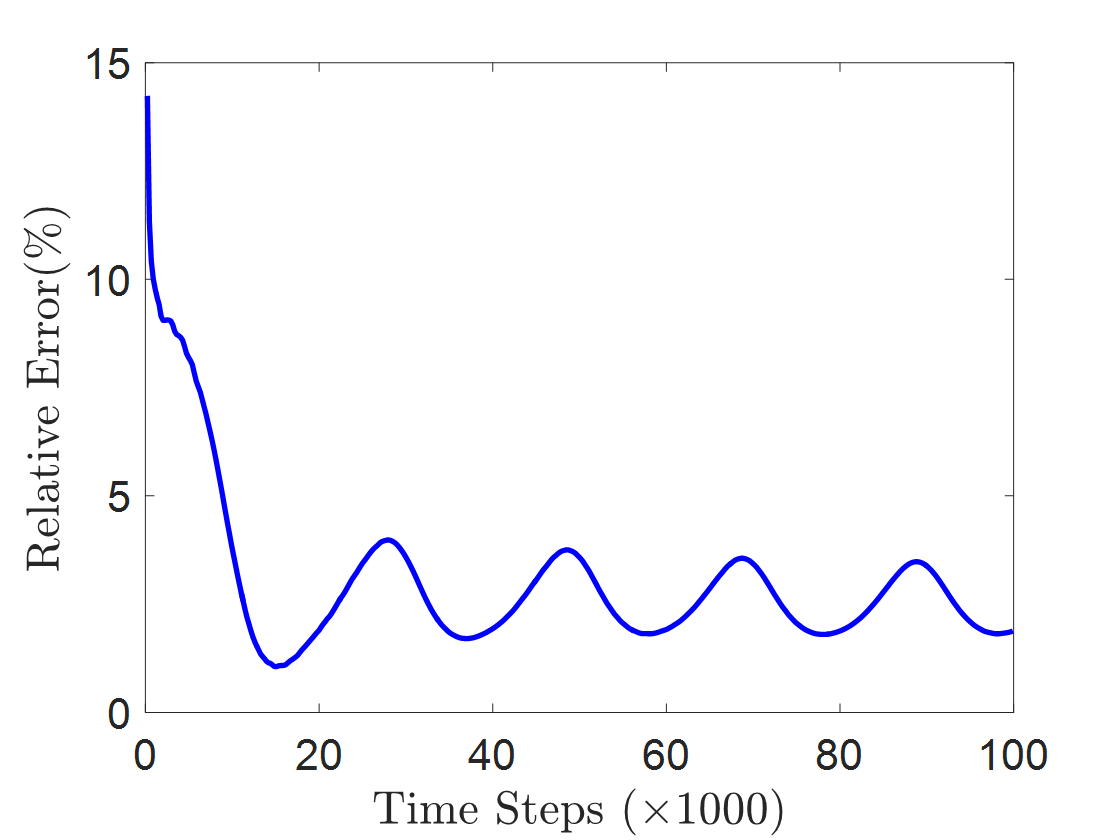}
\caption{The relative error of the ellipsoid rotation angle $\theta$ calculated based on the fluid solid coupling at $Re=0.36$. The fluid grid was $60\times 60\times 30$.}
\label{fig:JefferyRelativeError}
\end{figure}

To validate our IBM implementation, we considered the trajectory of a rigid ellipsoid in a shear flow in Stokes flow regime, commonly referred to as the Jeffery's orbit\cite{jeffery1922motion}, where the rotation angle of the ellipsoid, $\theta$, satisfies
\begin{equation}\label{eqn:jefferyOrbit}
tan\theta = \frac{b}{a}tan\frac{ab\gamma t}{a^2+b^2},
\end{equation}
where $\theta$ is the angle formed by the major axis of the ellipsoid and the shear flow direction, $a, b$ are the major and minor semi-axis lengths, $\gamma$ is the shear rate, and $t$ is the time. A 2D illustration of the ellipsoid is shown in Fig.\ref{fig:ellipsoidInShear}--the third semi-axis length is assumed to be the same as $b$.
An ellipsoid with semi-axis lengths $a=6, b=c=4.5$ was placed at the center of a channel with height $H=60$. The top and bottom velocities were set to $U=0.01$ and a viscosity of $\nu=1/6$ was prescribed. These parameters give a shear rate $\gamma=0.00033$ and a Reynolds number of $Re=aU/\nu=0.36$ to approximate Stokes flow. The IBM numerical result for the orientation angle $\theta$ and the analytical solution given by Eqn.(\ref{eqn:jefferyOrbit}) are plotted in Fig.\ref{fig:JefferyOrbitValidation}. The agreement is good, validating the FSI coupling. Shown in Fig.\ref{fig:JefferyRelativeError} is the relative error, defined as $\dfrac{|\theta_{sim}-\theta_{the}|}{\theta_{the}}$ where $\theta_{sim}$ is the simulation data and $\theta_{the}$ is the theoretical data from Eqn.\ref{eqn:jefferyOrbit}. The relative error is seen to oscillate with the period of the ellipsoid and is, on average, constant at about 2.62\%. 
\begin{figure}[H]
\centering
\includegraphics[width=0.6\textwidth]{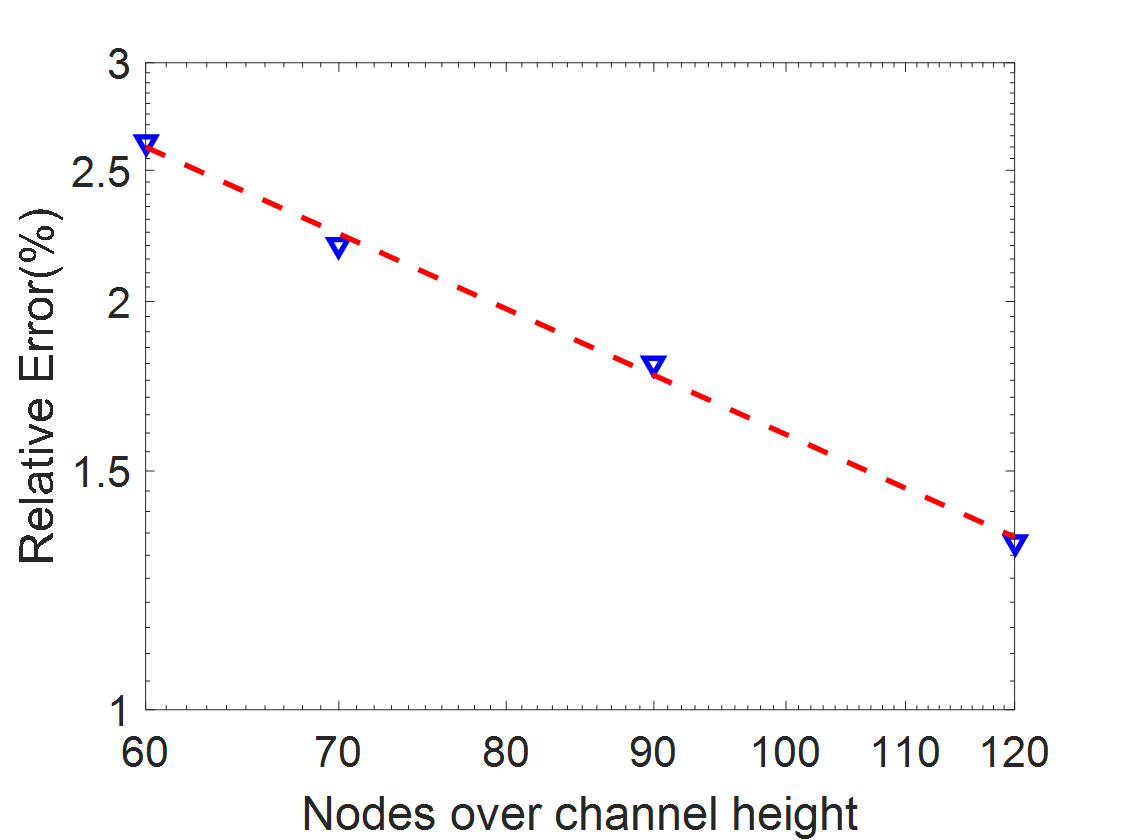}
\caption{The convergence for the fluid solid coupling at $Re=0.36$ with different grid resolutions. The convergence rate is nearly linear to the grid spacing, as indicated by the dashed line. The relative error $\epsilon \propto N^{-0.96}$ where $N$ is the grid resolution over channel height $H$.}
\label{fig:JefferyConvergence}
\end{figure}
As expected, the relative error $\epsilon$ is a power function of the fluid grid resolution $N$--Fig. \ref{fig:JefferyConvergence} shows that the relative error scales as the grid spacing following $ \epsilon \propto N^{-0.96}$. 

\subsection{Validation II: Red blood cell stretching test }\label{rst:validationRBCstretch}
\begin{figure}[H]
\begin{subfigure}{.4\textwidth}
  \centering
  \includegraphics[width=\linewidth]{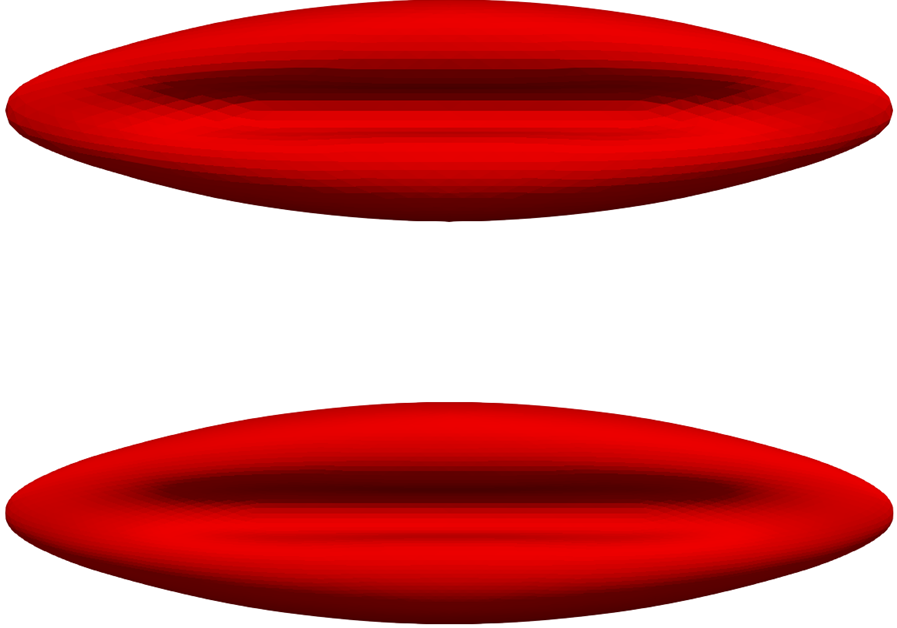}
  \caption{}
  \label{fig:200pNRBC}
\end{subfigure}%
\begin{subfigure}{.6\textwidth}
\centering
\includegraphics[width=\textwidth]{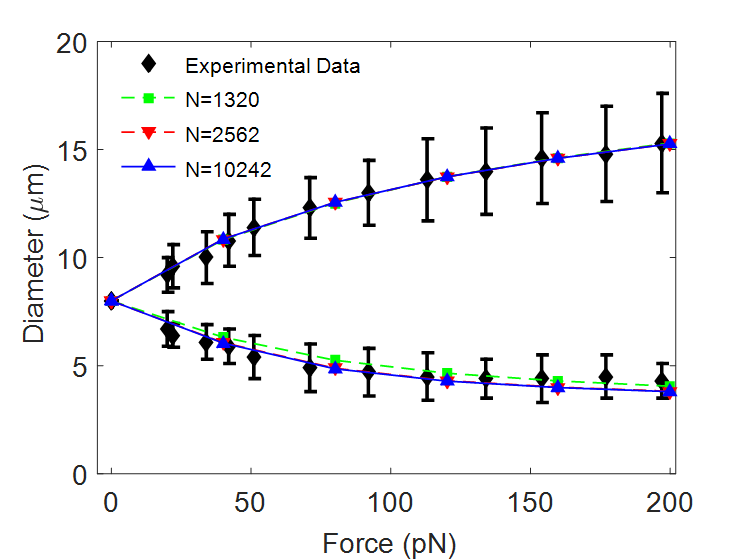}
\caption{}
\label{fig:rbcStretchingCurve}
\end{subfigure}%
\caption{(a) The deformed shape of a red blood cell under 200 pN using 2562 membrane particles (Top) and 10242 membrane particles (Bottom). (b) Comparison between the simulation results and optical tweezer experimental results\cite{dao2006molecularly} for the force diameter curve for red blood cells under different loadings. Simulation results were presented with three different membrane mesh size, e.g, N=1320, 2562, and 10242, respectively.}
\label{fig:rbcStretching}
\end{figure}

The RBC model was also validated and compared with optical tweezer stretching results under different loadings. The cell diameter without deformation was 8 $\mu m$, bending stiffness $k_b=2.3\times10^{-19} J$, $k_g=k_l=2.1\times10^{-4} N/m$, $ k_v=2.2 N/m^2$. 
The stretching constant $k_BT/p $ depended on the number of discretized membrane particles, e.g., $k_BT/p=1\times10^{-12} N$ for 1320 surface particles, $k_BT/p=7.14\times10^{-13} N$ for 2562 surface particles, $k_BT/p=3.57\times10^{-13} N$ for 10242 surface particles. However, the shear modulus $\mu_0$ was kept constant with $\mu_0=6$ $\mu N/m$ for all the meshes. A pair of forces was applied to two membrane patches on each side, occupying about 5\% of the total number of particles. The applied force was shared uniformly among these particles. Velocity verlet algorithm was used to integrate the Newton's equation. Viscous damping was used to stabilize the system. The deformation of RBCs under stretching (Fig.\ref {fig:200pNRBC}), particularly, the axial and transversal diameter (Fig.\ref{fig:rbcStretchingCurve}), were recorded and shown in Fig.\ref {fig:rbcStretching}. In Fig.\ref{fig:rbcStretchingCurve}, the top curve was the axial diameter, and the bottom curve was the transversal diameter. Both curves agreed well with experimental optical tweezer stretching data\cite{dao2006molecularly} for three different mesh sizes, validating the coarse grained molecular dynamics cell model. 

\subsection{Validation III: Effective blood viscosity}\label{rst:bloodViscosity}
\begin{figure}[H]
\begin{subfigure}{\textwidth}
  \centering
  \includegraphics[width=\linewidth]{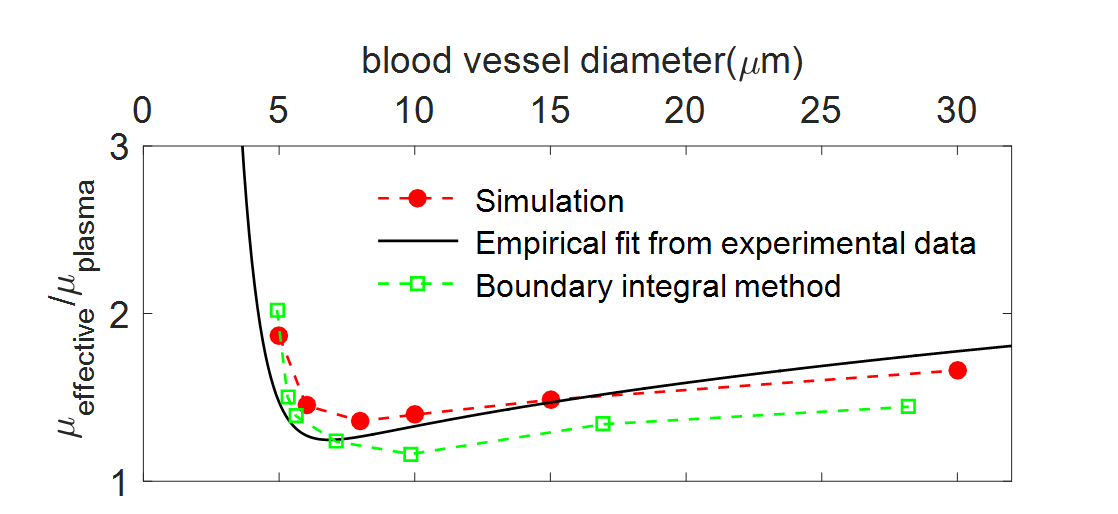}
  \caption{}
  \label{fig:bloodViscosity}
\end{subfigure}%
\hfill
\begin{subfigure}{\textwidth}
\centering
\includegraphics[width=0.85\textwidth]{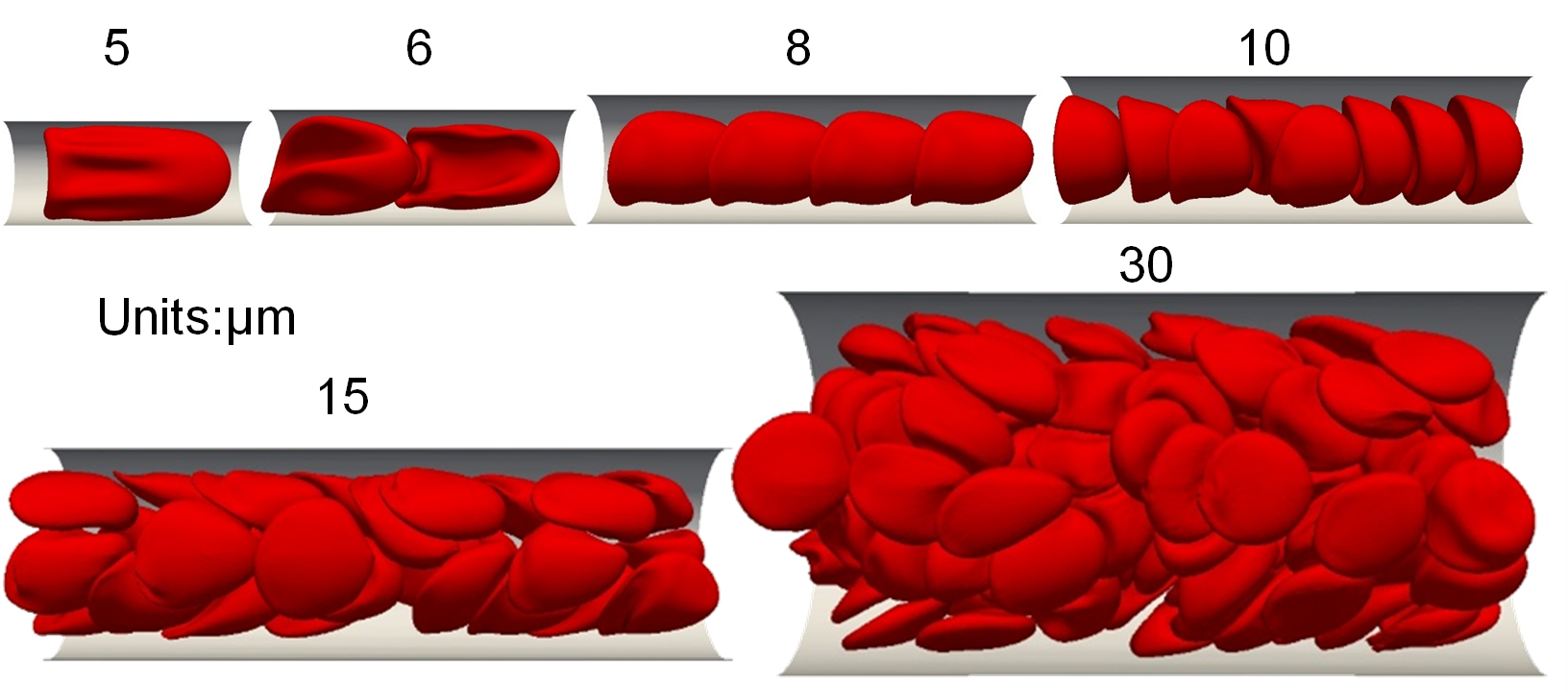}
\caption{}
\label{fig:bloodInTube}
\end{subfigure}%
\caption{(a). The normalized effective viscosity for blood flowing through tubes with different diameters. The data by boundary integral method was from Ref.\cite{freund2014numerical}, the empirical fit curve was from Ref.\cite{pries1992blood}.  (b) The snapshots for the red blood cells flowing through tubes with different diameters, showing different red blood cell shapes under flow. The hematocrit was 30\%.}
\label{fig:effectiveViscosity}
\end{figure}

Blood viscosity changes as it flows through different tube diameters and hematocrits. The viscosity decreases as the tube diameter in the range of 10$\sim$ 300 $\mu m$ due to the Fahraus-Lindqvist effect\cite{faahraeus1931viscosity}. The effective viscosity of blood in tube flow was studied and compared with experimental results\cite{pries1992blood}. The hematocrit defined as the actual volume ratio between cells and the fluid was 30\%. The tube diameter was ranging from 5 $\mu m $ to 30 $\mu m$. The fluid mesh size and the number of cell membrane nodes were 100 nm and 10242 for 5 and 6 $\mu m$ tube, 200 nm and 2562 for 8, 10, 15 $\mu m$, 333 nm and 2562 for 30 $\mu m$. The fluid mesh size was selected in a way such that it can resolve the fluid flow between cells and the tube wall. The mechanical properties were exact the same, as shown in section \ref{rst:validationRBCstretch}. A pressure gradient was applied to drive the flow. The Reynold's number defined with cell diameter and average flow speed was around 0.017 $\ll$ 1.  Periodic boundary conditions were applied at the inlet and outlet. The simulation ran for enough time to reach quasi-steady state, e.g., the volume rate was steady. The effective viscosity was normalized with respect to the plasma viscosity, and shown in Fig.\ref{fig:effectiveViscosity}. The simulation data clearly showed a decreasing viscosity as the tube diameter decreased from 30 to 8 $\mu m$. The viscosity increased as the tube diameter decreased further, which was resulted from the physical membrane contact with the wall. The viscosity curve agreed well with the empirical fitting from Ref.\cite{pries1992blood}, which validated that the model can capture the collective blood cell behavior under flow. The effective viscosity simulated by boundary integral method\cite{freund2014numerical} was also provided for comparison. Both methods can predict the blood viscosity reasonably well. 

\subsection{Scalability: Parallel performance}\label{rst:performance}
\begin{figure}[H]
\begin{subfigure}{0.5\textwidth}
\centering
\includegraphics[width=\textwidth]{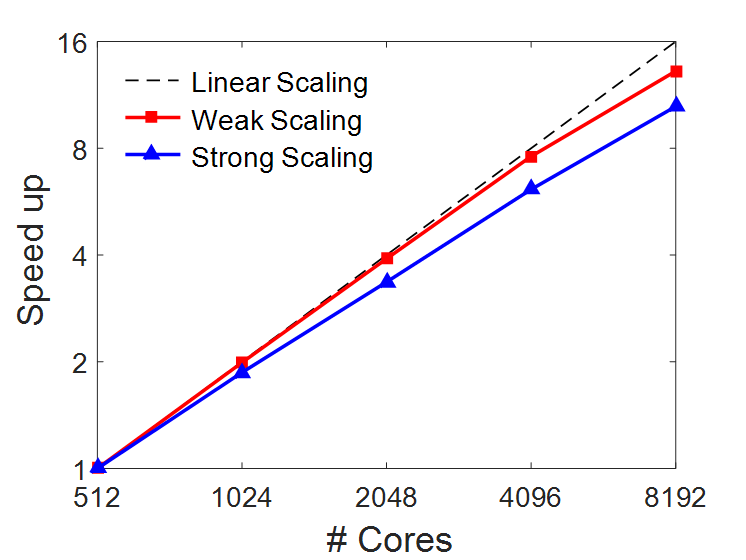}
\caption{Speed up of the simulation}
\label{fig:speedUp}
\end{subfigure}
\begin{subfigure}{0.5\textwidth}
  \centering
  \includegraphics[width=\linewidth]{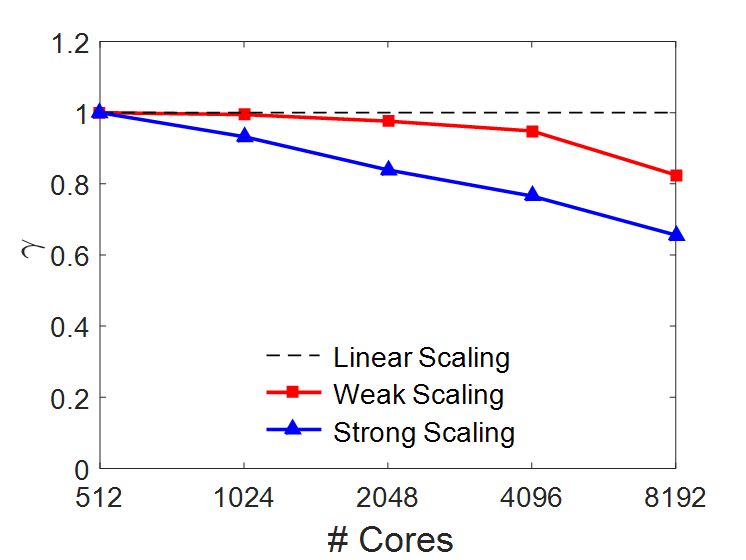}
  \caption{Scaling efficiency}
  \label{fig:scalingEfficiency}
\end{subfigure}
\caption{(a) The speed up of the simulation for strong and weak scaling cases. (b) The strong and  weak scaling efficiency of the code based on the immersed boundary method. The dashed lines showed the ideal linear scaling performance. All the simulations were performed on the IBM Blue Gene/Q system from Argonne National Laboratory. }
\label{fig:parallelPerformance}
\end{figure}

The parallel performance of the code was examined using a benchmark simulation of red blood cells flowing in a rectangular box. The fluid domain was discretized into a $360\times 360 \times 720$ grid in x,y,z directions with one inlet and one outlet and $4$ nonslip side walls. The fluid flow was driven by a body force of $1.34\times 10^{-7}$ with periodic inlet and outlet boundary conditions in the z direction. The solid phase consisted of $15,680$ red blood cells, discretized into $20,697,600$ particles connected by $61,998,720$ bonds, $41,332,480$ angles, and $61,998,720$ dihedrals in total. The hematocrit was about 40.8\%. The cells were initially uniformly distributed in the flow domain. The simulations were executed on the IBM Blue Gene/Q system from Argonne National Laboratory.  A ghost layer with thickness of three fluid lattice spacings was employed for inter-processor communication within the fluid solver. The ghost cutoff distance for \textit{LAMMPS} was set to 1.5 times the LBM lattice spacing. The CPU time for 100 time steps was recorded. Both strong scaling and weak scaling were considered. For weak scaling cases, the fluid domain size increased from $360\times 360\times 45$ for 512 cores to  $360\times 360 \times 720$  for 8192 cores. The parallelization performance was analyzed using two parameters: speed up ($SU$) and scaling efficiency ($\gamma$). The speed up and scaling efficiency were calculated as $
SU(N)=\dfrac{t_0^s}{t_N^s}, \gamma(N)=\dfrac{t_0^sN_0}{t_N^sN}$ for strong scaling cases, and $
SU(N)=\dfrac{t_0^wN_0}{t_N^wN}, \gamma(N)=\dfrac{t_0^w}{t_N^w}$ for weak scaling cases, where $t_0$ was the reference simulation time for $N_0$ cores, $t_N$ was the simulation time for $N$ cores. The subscript $s, w$ denoted strong and weak scaling cases, respectively. Here $N_0$ was taken as 512. The speed up of  the simulation time as well as scaling efficiency were shown in Fig.\ref{fig:parallelPerformance}. It showed that, as the number of cores increased from 512 to 8192, the speed up increased from 1 to 10.5 for strong scaling case, from 1 to 13.2 for weak scaling cases, see Fig.\ref{fig:speedUp}. Similarly, the scaling efficiency dropped from 1 to 0.656 for strong scaling case, and from 1 to 0.824 for weak scaling case, see Fig.\ref{fig:scalingEfficiency}. The dashed lines showed the ideal cases with no overhead and communication cost. It demonstrated that the combination of \textit{Palabos} and  \textit{LAMMPS} using the immersed boundary method can achieve high performance among thousands of processors. 

The scaling efficiency was found to depend strongly on the computer hardware. For example, less parallel efficiency was obtained on another computer with Intel Xeon E5630 2.53 GHz processors (16 cores in total) compared with 2 AMD Opteron 8-core 6128 processors with a clock speed of 2.0 GHz. While we did not study architecture dependence in more detail here, these results suggest that further work is needed to assess architecture-dependent performance issues. Finally, we note that the ghost layer thickness also influenced the performance of the IBM code. In the present study we used 3 layers of lattice points, while other studies have employed linear interpolation kernels that only require a single layer of lattice points for the communication layer\cite{mountrakis2015parallel}. 

\begin{figure}[H]
\centering
\includegraphics[width=0.6\textwidth]{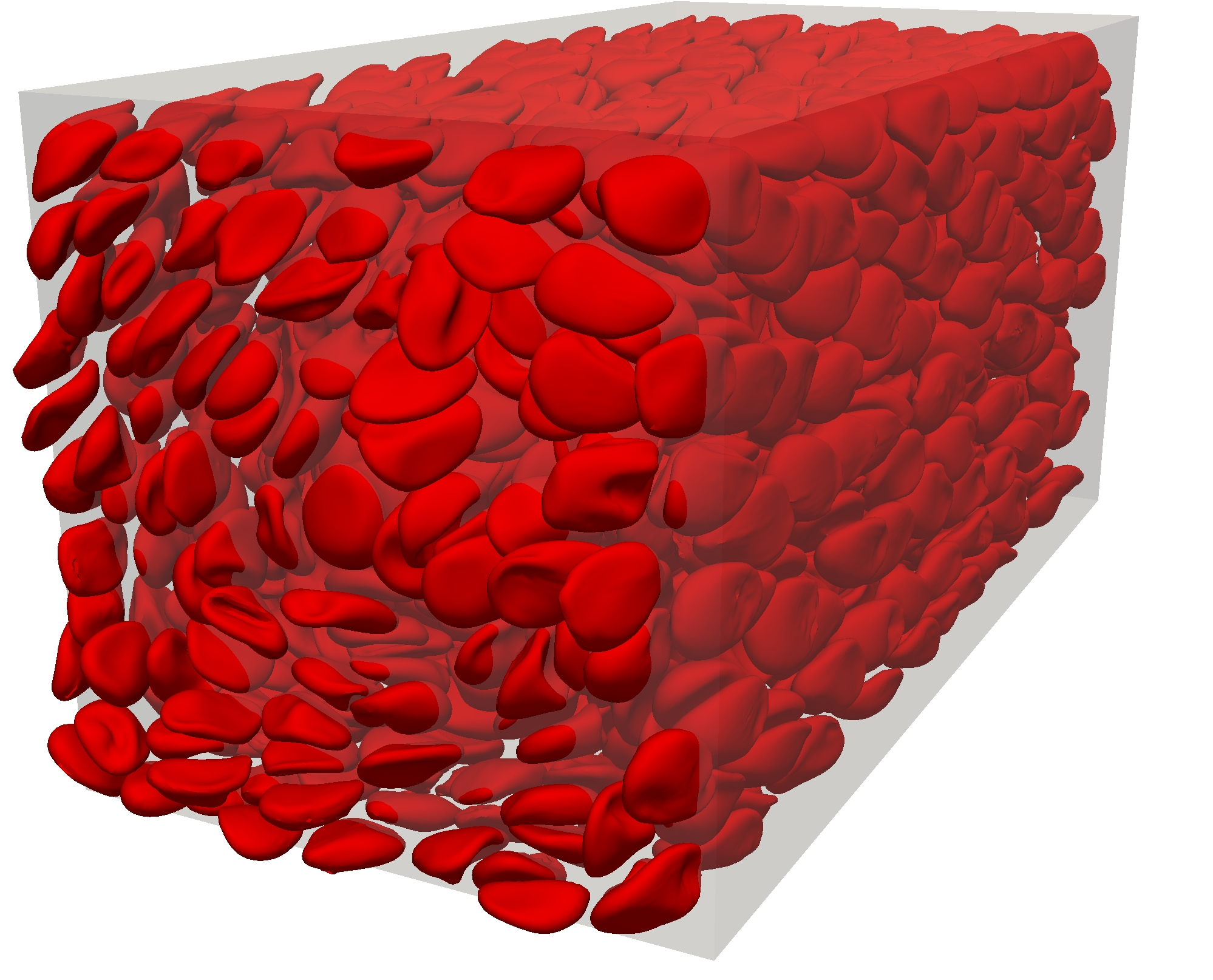}
\caption{A snapshot of the 1,470 blood cells moving in a microfluidic channel with a size of $60\times 60\times 120 \mu m$. The fluid domain was discretized into 11,664,000 lattices, the cells were discretized into 1,940,400 membrane nodes. The hematocrit level was 30\%.}
\label{fig:Ht30BloodChannel}
\end{figure}

The extra computational time induced by the coupling can be analyzed by GNU profiler tools. We considered the blood flow in a microfluidic channel with size of $60\times 60\times 120 \mu m$. $1,470$ cells were placed in the channel, resulting a hematocrit of 31\% in the flow.  In total there were $1,940,400$ cell membrane particles and $11,664,000$ fluid lattices, as shown in Fig.\ref{fig:Ht30BloodChannel}. On average, the extra computational time required for the information exchange between the fluid and solid solvers was about 34\%, e.g., time spent on the two coupling functions: interpolateVelocity3D and spreadForce3D. The computational time for the two coupling functions expressed in terms of the percentage over the total CPU time is shown in Table \ref{tab:timeCoupling}. The function interpolateVelocity3D cost about 9\% on average, while function spreadForce3D cost about 25\%. The percentage for both functions decreased as more cores were used. This was due to the increased communication cost among many processors. As expected, the coupling time would increase as the hematocrit level, as the interpolation and spreading was done for each cell membrane particle. 

\begin{table}[H]
\centering
\begin{tabular}{|l|l|l|l|l|l|l|}
\hline
\multirow{2}{*}{ID} & \multirow{2}{*} {Functions} & \multicolumn{5}{|c|}{CPU time (\%)} \\\cline{3-7} 
 & & p1 & p2 & p4 & p8 &p16\\
\hline 
$t_1$  & interpolateVelocity3D & 10.12 & 10.86 & 10.07 & 8.72 & 7.12\\
\hline
$t_2 $ & spreadForce3D   	  & 28.25   & 27.27  & 26.45   & 23.75   & 18.23\\
\hline
$t_1+t_2$   &                  & 38.37 & 38.12 & 36.51 & 32.46 & 25.35 \\
\hline
\end{tabular}
\caption{The percentage of CPU time for the coupling functions.}
\label{tab:timeCoupling}
\end{table}

\subsection{Case Study: Transport of flexible filaments in flowing red blood cell suspensions}\label{rst:app}
Next, the validated IBM code was applied to a problem relevant to drug carrier delivery in microcirculation flows to demonstrate its capabilities in a more complex setting. This is a multi-physics modeling problem as it involves fluid flow, large cell deformations, and polymer transport\cite{sinha2016shape,vahid2015microparticle,muller2014margination,vahid2015microparticle,tan2012influence}. Of particular interest is the observation that particles with different sizes, shapes, or flexibility can exhibit distinct transport properties in the blood stream. For example, long flexible filaments persist in the circulation up to one week after intravenous injection in rodent models. This is about ten times longer than for spherical counterparts\cite{geng2007shape}. The flexibility and filament length may influence transport and contribute to the anomalously long circulation time.

\begin{figure}[H]
\centering
\begin{subfigure}{.6\textwidth}
  \centering
  \includegraphics[width=\linewidth]{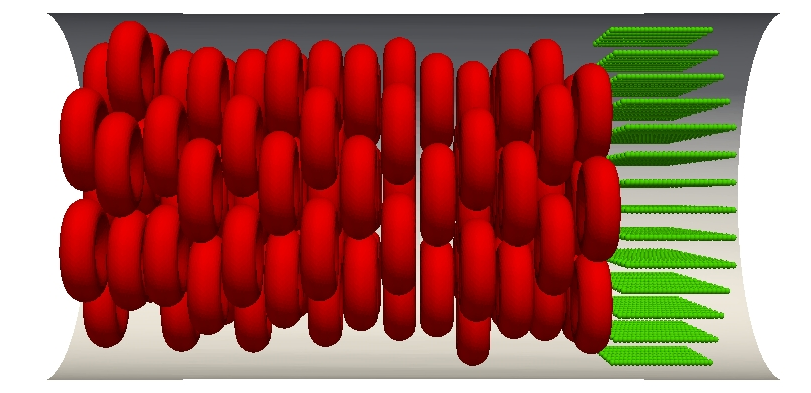}
  \caption{Front view}
  \label{fig:frontT0}
\end{subfigure}%
\begin{subfigure}{.4\textwidth}
  \centering
  \includegraphics[width=0.75\linewidth]{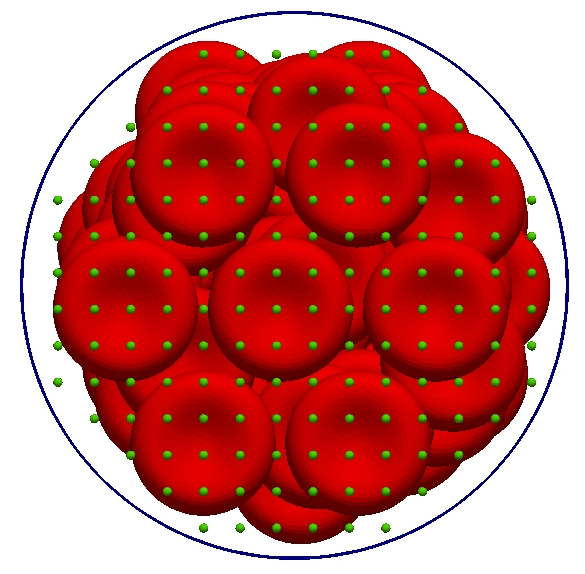}
  \caption{Side view}
  \label{fig:sideT0}
\end{subfigure}
\caption{Initial positions for the blood cells and filaments. The cells were randomly distributed in the blood vessel, while the filaments were uniformly distributed in one end of the vessel.}
\label{fig:initialSetup}
\end{figure}

We simulated filament transport in tube flow of red blood cell suspensions and analyzed filament migration, apparent size, spatial distribution, and dispersion rate. The fluid domain was represented by a cylindrical tube with diameter of $30\mu m$ and length $50 \mu m$, as shown in Fig.\ref{fig:initialSetup}. No-slip conditions were applied at the cylinder wall while periodic boundary conditions were applied at the inlet and outlet for both fluid and cells. The relaxation time was set to $\tau=1$ giving a dimensionless kinetic viscosity of $1/6$. The LBM lattice spacing, $dx$, was fixed at $0.33\mu m$, the time step was set to $dt=1.82\times10^{-8}s$. One hundred red blood cells were placed in the domain corresponding to a cell concentration of about 27\%, which is within the physiological range\cite{boyle1988microcircu}. Each cell was described by 1,320 surface nodes, 3,954 stretching bonds, 2,636 area segments, and 3,954 bending angles. A force density was applied to drive the flow such that the shear rate at the wall was $1000 s^{-1}$ without the red blood cells. The actual wall shear rate was less as the blood cell component reduces the flow rate. A total of 156 filaments were then introduced at one end of the cylinder as shown in Fig.\ref{fig:initialSetup}. Each filament was a polymer chain modeled as beads connected by stretching and bending springs. Different lengths ($2\mu m $ and $8\mu m$) and different bending stiffnesses ($4.1\times10^{-20}$ J, or about $10 k_BT $, and $4.1\times10^{-18}$ J, corresponding to about $1000 k_BT$) were used to study the size and stiffness effect on filament transport in red blood cell suspensions.

The parameters used in the simulations are listed in Table.\ref{tab:paraPtlCellMix}. The corresponding shear modulus for red blood cell membranes was $\mu_0=6\mu N/m$ based on Eqn.\ref{eqn:shearModulus}. 

\begin{table}[H]
\centering
\begin{tabular}{|l|l|l|l|l|}
\hline
\multicolumn{5}{|l|}{Red blood cells}\\
\hline
$k_BT/p (N)$&$k_b (J)$ &$k_g (N/m)$ &$k_l (N/m)$ &$ k_v (N/m^2)$\\
\hline 
$1\times10^{-12}$  & $2.3\times10^{-19}$ &  $2.1\times10^{-4}$& $2.1\times10^{-4} $& $2.2$ \\
\hline
\multicolumn{5}{|l|}{Filaments}  \\
\cline{1-5}
$k_BT/p (N)$ &$k_p (J)$&$k_b (J)$&$l_0 (\mu m)$ &$\theta_0 (^{\circ})$\\
\cline{1-5}
$1.36\times10^{-12}$ & $1.7\times10^{-12}$ & $[4.1,410]\times10^{-20} $ &0.33&$180$  \\
\cline{1-5}
\multicolumn{4}{|l|}{Morse potential}  \\
\cline{1-4}
$D_0(J)$&$\alpha$&$r_0(\mu m)$&$r_c(\mu m)$\\
\cline{1-4}
$1.2\times10^{-18}$&$0.5$&$0.66$&$0.66$\\
\cline{1-4}
\end{tabular}
\caption{Simulation parameters used for the particle transport in blood flow. The shear modulus for the cell $\mu_0=6\mu N/m$. The stretching and bending parameters for filaments were selected based on \cite{baschnagel2016semiflexible,ness2017nonmonotonic}. The filaments were assumed to be straight in equilibrium, and the equilibrium length was chosen due to the resolution of the fluid lattice. The Morse potential was used to avoid the filament-cell and cell-cell overlapping.}
\label{tab:paraPtlCellMix}
\end{table}

\begin{figure}[H]
\centering
\begin{subfigure}{.5\textwidth}
  \centering
  \includegraphics[width=\linewidth]{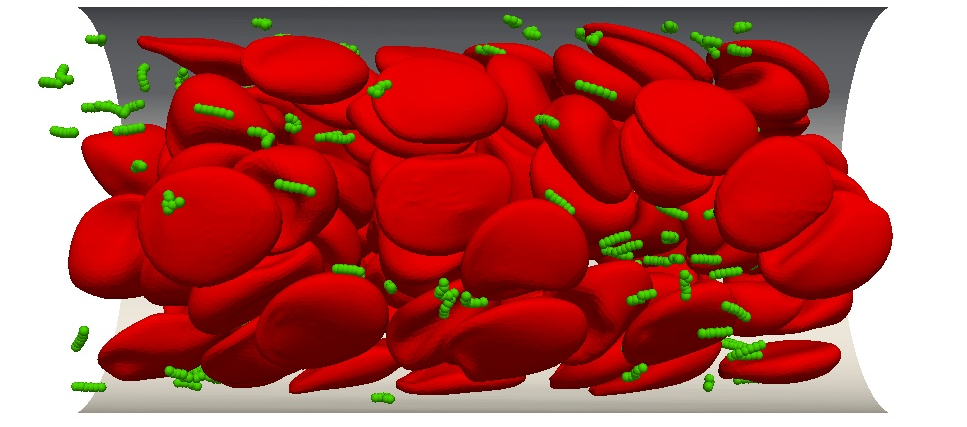}
  \caption{$2\mu m, k_b=10k_BT$}
  \label{fig:10kBT2um0400}
\end{subfigure}%
\begin{subfigure}{.5\textwidth}
  \centering
  \includegraphics[width=\linewidth]{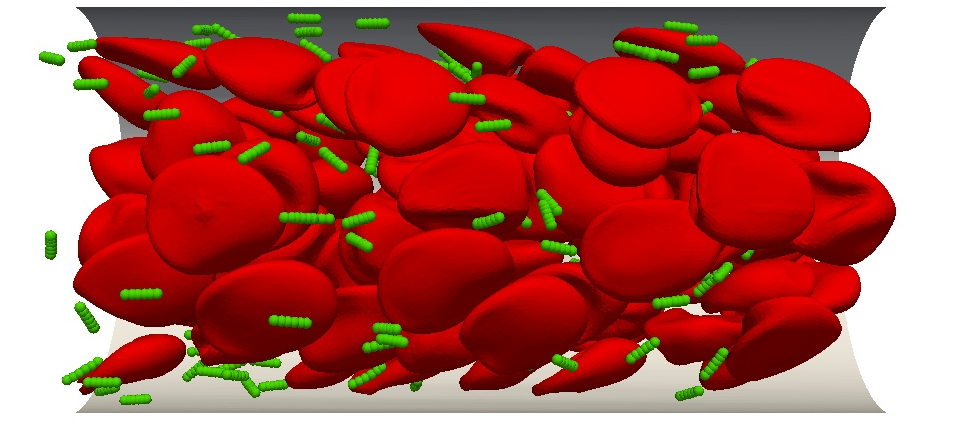}
  \caption{$2\mu m, k_b=1000k_BT$}
  \label{fig:1e3kBT2um0400}
\end{subfigure}
\begin{subfigure}{.5\textwidth}
  \centering
  \includegraphics[width=\linewidth]{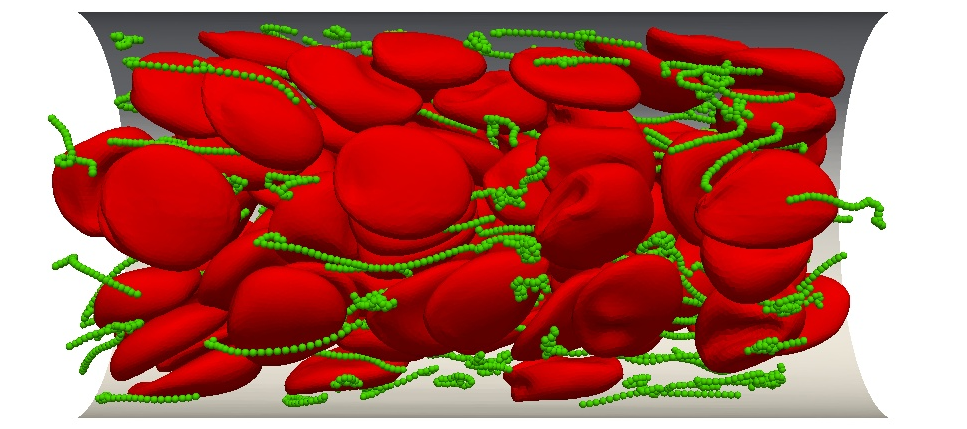}
  \caption{$8\mu m, k_b=10k_BT$}
  \label{fig:10kBT8um0400}
\end{subfigure}%
\begin{subfigure}{.5\textwidth}
  \centering
  \includegraphics[width=\linewidth]{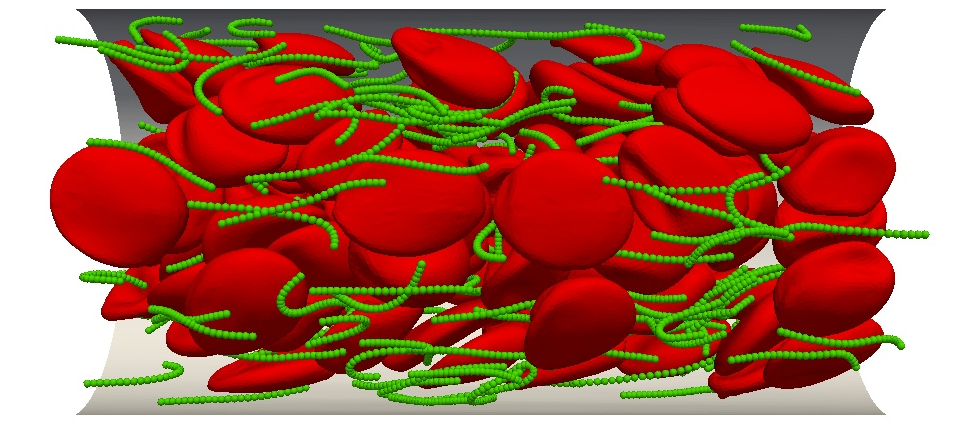}
  \caption{$8\mu m, k_b=1000k_BT$}
  \label{fig:1e3kBT8um0400}
\end{subfigure}
\caption{Snapshots of the simulations of filaments (green) mixing with blood cell suspensions (red) at $8\times10^6$ time steps. 2$\mu m$ filaments with bending stiffness $10 k_BT$ (a) and $1000 k_BT$ (b); 8$\mu m$ filaments with bending stiffness $10 k_BT$ (c) and $1000 k_BT$ (d). Half of the vessel wall was also shown in the figure.}
\label{fig:cellParticleMix}
\end{figure}
Simulation snapshots at $8\times10^{6}$ time steps are shown in Fig.\ref{fig:cellParticleMix} for cases with different filament size and flexibility. Different configurations are observed for filaments with different stiffnesses. The stiffer filaments were generally straight when the length was $2\mu m$, and exhibited a small amount of curvature for $8\mu m$, see Fig.\ref{fig:1e3kBT2um0400} and \ref{fig:1e3kBT8um0400}. By contrast, the more flexible filaments exhibited highly bent or coiled shapes, particularly for the $8\mu m$ filaments, see Fig.\ref{fig:10kBT2um0400} and \ref{fig:10kBT8um0400}. The apparent filament size, defined as the maximum extent for the filaments, is shown in Table.\ref{tab:filamentsLength}. The apparent size for stiffer filaments was very close to the actual contour length, while the apparent size for flexible filaments was smaller than the actual contour length, due to the bending and coiling of the filaments.
\begin{table}[h!]
\centering
\begin{tabular}{|l|l|l|l|l|}
\hline
contour size($\mu m$) & \multicolumn{2}{|c|}{$1.66 $}&\multicolumn{2}{|c|}{$7.92 $}\\
\hline
stiffness($k_b/k_BT$) &$10$ &$1000$ &$10$ & $1000$\\
\hline 
apparent size($\mu m$) & \small{$1.57\pm 0.31$}  & \small{$1.67\pm 0.01$} &  \small{$5.76\pm 2.4$}& \small{$7.89\pm 1.92 $} \\
\hline
\footnotesize{dispersion rate($\times 10^{-11}m^2/s$)}&$9.76$&$8.64  $&$6.26  $&$7.26  $\\
\hline
\end{tabular}
\caption{The apparent size and dispersion rate of the filaments in the blood flow with cells. The apparent size of the filaments were smaller than the actual contour length due to the bending and coiling. The dispersion rate was in the order of $10^{-11}m^2/s$, which was about 2 orders-of-magnitude larger than thermal diffusion for microparticles.}
\label{tab:filamentsLength}
\end{table}

\begin{figure}[H]
\centering
\begin{subfigure}{.5\textwidth}
  \centering
  \includegraphics[width=\linewidth]{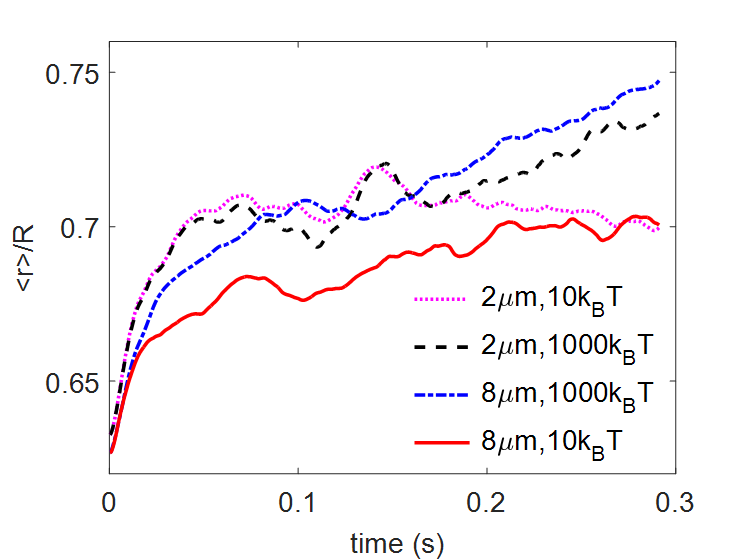}
  \caption{Normalized radial position}
  \label{fig:radialPosition}
\end{subfigure}%
\begin{subfigure}{.5\textwidth}
  \centering
  \includegraphics[width=\linewidth]{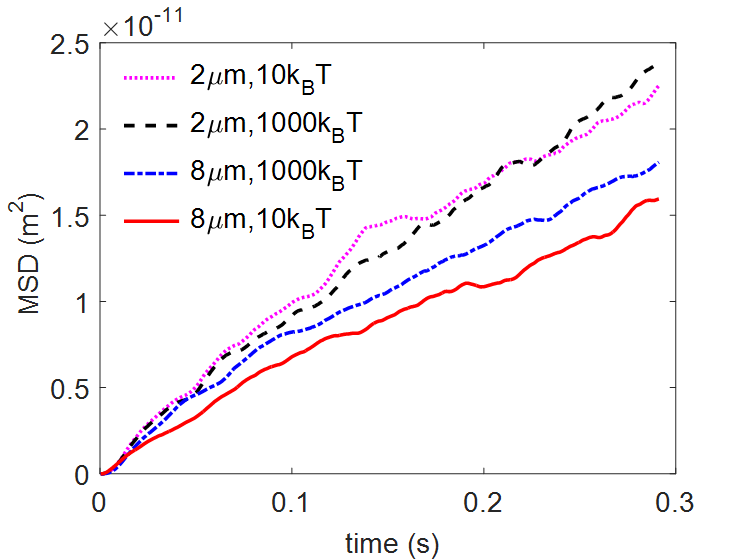}
  \caption{Mean Square Displacement}
  \label{fig:MSD}
\end{subfigure}
\caption{(a) The averaged filament center position normalized with the vessel radius $R=15\mu m$. Stiffer filaments showed better margination properties than flexible filaments. (b) The mean square displacement for the filaments transported with the blood cells.}
\label{fig:averageMSD}
\end{figure}

The average radial position $\left<r\right>$ normalized to the vessel radius $R$ and the mean square displacement (MSD) for the filaments are plotted in Fig.\ref{fig:radialPosition} and Fig.\ref{fig:MSD}. Filament migration towards the vessel wall was observed for all filament types, as indicated by the increasing values of $\left<r\right>/R$. The averaged mean radial positions for flexible filaments ($10k_BT$) was stabilized around $0.7R$ for both sizes. For stiffer filaments ($1000k_BT$), the $\left<r\right>/R$ kept increasing within the simulation time, showing better margination properties than flexible ones. This agrees well with previous findings for deformable particles\cite{muller2016understanding}. Less margination for flexible ones lead to low binding to the vessel wall and high concentration in the blood stream, which can also explain the long circulation time for long flexible filaments in rodents\cite{geng2007shape}. The dispersion rate, defined as the slope of the MSD curve, is shown in Table \ref{tab:filamentsLength}. The dispersion rates are all of the order of $10^{-11} m^2/s$, which is about 2 orders-of-magnitude larger than the background thermal diffusion expected for particles of this size. This finding is also consistent with other observations for small particles in blood suspensions, such as platelets\cite{zhao2012shear,vahid2014platelet,crowl2010computational}, particles\cite{vahid2015microparticle,muller2016understanding} and experimental measurements for microparticles\cite{saadatmand2011fluid}.  

\begin{figure}[H]
\centering
\begin{subfigure}{.5\textwidth}
  \centering
  \includegraphics[width=\linewidth]{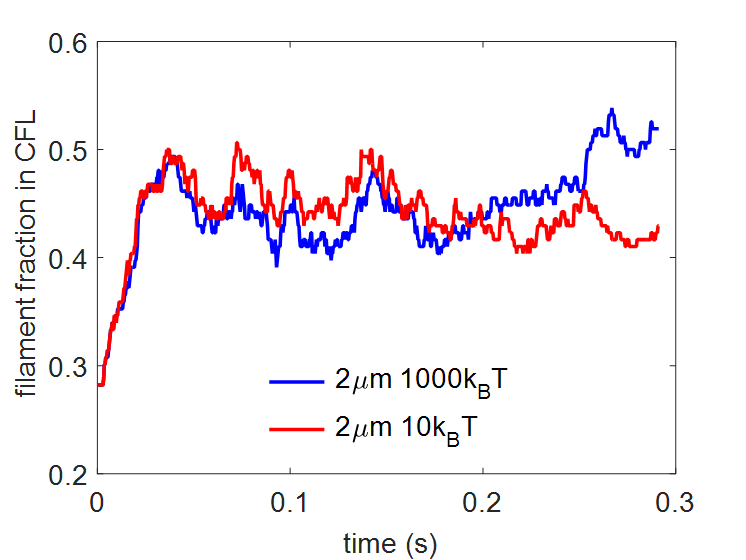}
  %\caption{$2\mu m$}
  \label{fig:2umFraction}
\end{subfigure}%
\begin{subfigure}{.5\textwidth}
  \centering
  \includegraphics[width=\linewidth]{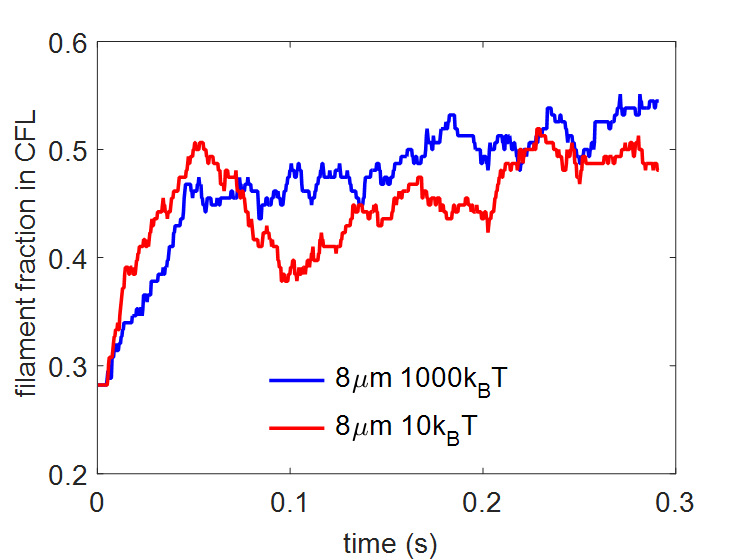}
 % \caption{$8\mu m$}
  \label{fig:8umFraction}
\end{subfigure}
\caption{The fraction of the filaments in the cell free layer of the blood vessel. The cell free layer was defined as a 3$\mu m$ thickness layer next to the vessel wall. (a) $2\mu m$; (b)$8\mu m$.}
\label{fig:filamentFraction}
\end{figure}

The fraction of filaments present in the $3\mu m$ cell-free layer (CFL) near the vessel wall is shown in Fig.\ref{fig:filamentFraction}. Only the data in the last 0.15 s were selected for analysis due to the transient effect in the beginning. The fraction of the 2$\mu m$ filaments in the CFL increases approximately from 28\% (the baseline value assuming a uniform distribution) to $ 43.3\%\pm 1.7\%$ for soft filaments, and to $46.1\%\pm 3.4\%$ for stiff filaments.  However, the fraction of 8$\mu m$ filaments in the CFL reached $47.5\%\pm 2.4\%$ for the flexible ones, and  $51.1\%\pm 1.9\%$  for stiffer ones.  These results suggest that longer and stiffer filaments in blood cell suspensions marginated quickly than the shorter and softer ones. The margination for rigid filaments also required less traveling distance in the flow direction, as the effective viscosity was increased for the mixture of rigid filaments in blood cell suspensions.  

\section{Conclusions}\label{con}
We introduced an efficient implementation of the Immersed Boundary Method by coupling a Lattice Boltzmann fluid solver with a particle-based solid solver using open source codes, namely \textit{Palabos} and \textit{LAMMPS}, respectively. The coupling was achieved by a simple interface that only requires a few hundred lines of code, dramatically reducing software development time, increasing robustness and facilitating transferability. The coupling was demonstrated to be computationally efficient. The extra computing time depends on the solid fraction, e.g., for a hematocrit of 30\%, it required only about a 34\% increase in the computational cost beyond what was required for executing fluid and solid solvers.  The IBM code was also shown to scale linearly from 512 to 8192 cores for both strong and weak scaling cases, demonstrating great parallelization efficiency in massively parallel environments.

The validated IBM code was used to analyze polymer filament transport in red blood cell flows; such filaments are of interest for drug delivery applications. Simulations demonstrated how filament flexibility reduce the effective size of filament, so that flexible filaments can traverse through the blood cell suspensions easier than stiffer ones. The stiffer filaments showed better margination properties than flexible ones, consistent with previous findings\cite{muller2016understanding}. These results highlight the complex physics that may underlie the long circulation time found for such particles in rodents \cite{geng2007shape}. 

\section{Acknowledgment}
This work was supported by NIH 1U01HL131053-01A1. We thank the high performance computing support from Gaea at Northern Illinois University and the Mira at the Argonne National Laboratory. 

\section*{References}

\bibliography{./references}

\end{document}